\documentclass[preprintnumbers,prd,showpacs,floatfix,superscriptaddress,nofootinbib,twocolumn, letterpaper]{revtex4-1}
\pdfoutput=1

\usepackage{longtable}
\usepackage{bm}
\usepackage{relsize}
\usepackage{amsfonts}
\usepackage{amsmath}
\usepackage{amssymb,epsf}
\usepackage{latexsym}
\usepackage{graphicx,epsfig}
\usepackage{amssymb}
\usepackage{float}
\usepackage{subfigure}
\usepackage{epstopdf}
\usepackage[colorlinks=true,citecolor=blue,linkcolor=blue,urlcolor=black]{hyperref}
\usepackage{dcolumn}
\usepackage{psfrag}
\usepackage{wrapfig}
\usepackage{makeidx}
\usepackage{epsf}
\usepackage{color}
\usepackage{multirow}
\usepackage{mathtools}
\usepackage{hyperref}
\hypersetup{colorlinks=true, linkcolor=blue, citecolor=green}
\usepackage{enumerate}
\usepackage{tikz}

\newcommand{\orcidicon}{%
	\begin{tikzpicture}
	\draw[lime, fill=lime] (0,0)
		circle [radius=0.16]
		node[white] {{\fontfamily{qag}\selectfont \tiny ID}};
	\draw[white, fill=white] (-0.0625,0.095)
		circle [radius=0.007];
	\end{tikzpicture}	\hspace{-2mm}
}
\newcommand\orcidRosa{{\href{https://orcid.org/0000-0003-4148-7372}{\orcidicon}}}
\newcommand\orcidMarques{{\href{https://orcid.org/0000-0001-7022-5502}{\orcidicon}}}
\newcommand\orcidBazeia{{\href{https://orcid.org/0000-0003-1335-3705}{\orcidicon}}}
\newcommand\orcidLobo{{\href{https://orcid.org/0000-0002-9388-8373}{\orcidicon}}}

\begin{document}

\title{Thick branes in the scalar-tensor representation of $f(R,T)$ gravity}

\author{Jo\~{a}o Lu\'{i}s Rosa\orcidRosa\!\!}
\email{joaoluis92@gmail.com}
\affiliation{Institute of Physics, University of Tartu, W. Ostwaldi 1, 50411 Tartu, Estonia}

\author{Matheus A. Marques\orcidMarques\!\!}
\email{marques@cbiotec.ufpb.br}
\affiliation{Departamento de Biotecnologia, Universidade Federal da Para\'\i ba, 58051-900 Jo\~ao Pessoa, PB, Brazil}
\affiliation{Departamento de F\'isica, Universidade Federal da Para\'iba, 58051-970, Jo\~ao Pessoa, PB, Brazil}

\author{Dionisio Bazeia\orcidBazeia\!\!}
\email{dbazeia@gmail.com}
\affiliation{Departamento de F\'isica, Universidade Federal da Para\'iba, 58051-970, Jo\~ao Pessoa, PB, Brazil}

\author{Francisco S. N. Lobo\orcidLobo\!\!}
\email{fslobo@fc.ul.pt}
\affiliation{Instituto de Astrof\'{i}sica e Ci\^{e}ncias do Espa\c{c}o, Faculdade de Ci\^encias da Universidade de Lisboa, Edif\'{i}cio C8, Campo Grande, P-1749-016, Lisbon, Portugal}
\affiliation{Departamento de F\'{i}sica, Faculdade de Ci\^{e}ncias, Universidade de Lisboa, Edifício C8, Campo Grande, PT1749-016 Lisbon, Portugal}

\date{\today}

\begin{abstract} 
Braneworld scenarios consider our observable universe as a brane embedded in a five-dimensional bulk. In this work, we consider thick braneworld systems in the recently proposed dynamically equivalent scalar-tensor representation of $f(R,T)$ gravity, where $R$ is the Ricci scalar and $T$ the trace of the stress-energy tensor. In the general $f\left(R,T\right)$ case we consider two different models: a brane model without matter fields where the geometry is supported solely by the gravitational fields, and a second model where matter is described by a scalar field with a potential. The particular cases for which the function $f\left(R,T\right)$ is separable in the forms $F\left(R\right)+T$ and $R+G\left(T\right)$, which give rise to scalar-tensor representations with a single auxiliary scalar field, are studied separately. The stability of the gravitational sector is investigated and the models are shown to be stable against small perturbations of the metric. Furthermore, we show that in the $f\left(R,T\right)$ model in the presence of an extra matter field, the shape of the graviton zero-mode develops internal structure under appropriate choices of the parameters of the model.
\end{abstract}


\maketitle

\section{Introduction}\label{sec:intro}

Braneworld scenarios consider our observable universe as a brane embedded in a 5D bulk.
They were first considered in Refs. \cite{Randall:1999vf,Goldberger:1999uk,DeWolfe:1999cp,Csaki:2000fc} and then in a diversity of scenarios engendering standard and modified gravity. More recently, thick braneworld systems were constructed in the proposed $f(R,T)$ gravity \cite{Harko:2011kv}, with $R$ the Ricci scalar and $T$ the trace of the stress-energy tensor; see, e.g., Refs. \cite{Bazeia:2015owa,Gu:2016nyo}. Analytic background solutions were obtained and the full linear perturbations were explored, especially the metric \cite{Bazeia:2015owa} and scalar perturbations \cite{Gu:2016nyo}. 
It was shown explicitly that under specific situations, the gravity sector of this new braneworld scenario is linearly stable \cite{Bazeia:2015owa}. Due to the rich internal structure of $f(R,T)$ gravity, more interesting background braneworld solutions as compared to general relativity coupled to a canonical scalar field were found \cite{Gu:2016nyo}. It was shown that there is no tachyon state in this model and only the massless tensor mode can be localized on the brane, which recovers the effective four-dimensional gravity.
Thick branes were re-examined in $f(R,T)$ gravity \cite{Rohman:2021vvv}, where the equation of motion for the scalar field and the $f (R,T)$ field equation for a conformally flat brane and Robertson-Walker brane were derived. The localization of the scalar field on the $f (R, T)$ thick brane was also explored, for a specific $f (R,T)$ form.

In Ref. \cite{Moraes:2015dee}, it was shown that the cosmological parameters obtained from 5D $f(R,T)$ gravity are in agreement with recent constraints from type Ia supernovae data, baryon acoustic oscillations and cosmic microwave background observations, favoring such an alternative description of the universe dynamics.
In this context, $f(R,T)$ gravity was also investigated in the so-called configurational entropy (CE) context \cite{Correa:2015qma}. It was shown, by means of this information-theoretical measure, that a stricter bound on the parameter of $f(R,T)$ brane models arises from the CE. It was further found that these bounds are characterized by a valley region in the CE profile, where the entropy is minimal. It was argued that the CE measure can open a new role and an important additional approach to select parameters in modified gravity.

The search of thick brane solutions is an extremely active branch of research and a plethora of scenarios have recently been extensively explored, in particular, within modified gravity. As far as we know, the first work on branes in $f(R)$ theory appeared in Ref. \cite{Afonso:2007gc} and more recently in  \cite{Zhong:2010ae,Bazeia:2013oha,Bazeia:2013uva,Bazeia:2014poa,Bazeia:2015zpa,Gu:2014ssa,daSilva:2017jbx,Gu:2018lub} with distinct motivations. Moreover, after the reviews on modified gravity \cite{DeFelice:2010aj,Nojiri:2010wj}, several works on thick branes on extended models have investigated issues of current interest; see, e.g., Refs. \cite{Guo:2018tpo,Cui:2020fiz,Rosa:2020uli,Moreira:2021xfe,Chen:2020zzs,Bazeia:2020zut,Bazeia:2020jma,Xie:2021ayr,Xiang:2020qrc,Moreira:2021vcf} and references therein. For instance, in Ref. \cite{Cui:2020fiz} the authors studied thick brane solutions in a 6D spacetime in modified $f(R)$ gravity; Ref. \cite{Guo:2018tpo} considered a modified mimetic model controlled by the presence of the scalar torsion;  5-dimensional braneworld scenarios were analyzed in the scalar-tensor representation of the generalized hybrid metric-Palatini gravitational theory \cite{Rosa:2020uli}; brane structure and gravitational resonances of thick branes generated by a mimetic scalar field in $f(R)$ gravity were studied in \cite{Chen:2020zzs}; 5-dimensional $f(T,B)$ teleparallel modified gravity braneworld scenarios, where asymptotically, the bulk geometry converges to an $AdS_5$ spacetime whose cosmological constant is produced by parameters that control torsion \cite{Moreira:2021xfe}; 
a novel approach was analyzed in \cite{Bazeia:2020zut} in modified gravity with Lagrange
multipliers, where the linear stability of the models were explored; the case of flat and bent branes with internal structure in mimetic gravity in the presence of two real scalar fields \cite{Xiang:2020qrc}; the study in \cite{Bazeia:2020jma} to show that including a Lagrange multiplier unveils an alternative approach to induce brane structure using a single scalar field, tracing out new avenues of research in braneworld scenarios, naturally leading to interesting results for the localization of matter fields in the brane; the investigation of thick branes generated by a scalar field in mimetic gravity theory \cite{Xie:2021ayr}, where the presence of two auxiliary superpotentials is considered to change the second order field equations into first order equations; and also, the study of fermion localization in branes in another scenario in teleparallel $f(T,B)$ gravity \cite{Moreira:2021vcf}. 

The recent results on thick braneworld solutions developed in Refs. \cite{Rosa:2020uli,Bazeia:2020jma} in modified gravity motivate us to further study the issue, searching for new scenarios and solutions. In particular, in this work we will focus in the recently proposed dynamically equivalent scalar-tensor representation of $f\left(R,T\right)$ gravity which clarifies the two extra scalar degrees of freedom of the theory via the introduction of two gravitational scalar fields \cite{Luis:2021xay}. The present work is outlined in the following manner: In Sec. \ref{sec:theogen}, we introduce the general model, the source matter distribution and the equations of motion in the case of a five-dimensional line element appropriated to study braneworld models in the presence of a single extra dimension of infinite extent. We also investigate the stability of the gravitational sector. In Sec. \ref{sec:solsgen}, we derive two solutions in the general $f\left(R,T\right)$ case, the first without matter and the second in the presence of a source matter field. We then move on to Sec. \ref{sec:other}, where two interesting new possibilities are considered, one with $f(R,T)=f(R)+ T$ and the other with $f(R,T)=R+f(T)$. These cases are of particular interest since they describe braneworld scenarios that fall within the class of models with stable gravitational sector \cite{Bazeia:2015owa} and may unveil a new route for the study of thick branes in modified gravity in the presence of the trace of the stress-energy tensor \cite{Harko:2011kv}. Finally, in Sec. \ref{sec:concl}, we conclude and discuss possible lines of future research in the subject.

\section{Action and field equations}\label{sec:theogen}
The action that describes $f\left(R,T\right)$ gravity \cite{Harko:2011kv} in $4+1$ dimensional gravity is given by
\begin{equation}\label{actiongeo}
S=\frac{1}{2\kappa^2}\int_\Omega\sqrt{-g}f\left(R,T\right)d^5x+S_m\left(g_{MN},\chi\right),
\end{equation}
where $\kappa^2=8\pi G_5$, $G_5$ is the 5-dimensional Newtonian constant, $\Omega$ is a 5-dimensional spacetime manifold on which we define a set of coordinates $x^M$, $g$ is the determinant of the metric $g_{MN}$, $f$ is an arbitrary function of the Ricci scalar $R=g^{MN}R_{MN}$, where $R_{MN}$ is the Ricci tensor, and the trace $T$ of the stress-energy tensor $T_{MN}$, $S_m$ is the matter action defined as $S_m=\int\mathcal L_m d^5x$, where $\mathcal L_m$ is the matter Lagrangian density considered minimally coupled to the metric $g_{MN}$, and $\chi$ collectively denotes the matter fields.

The modified field equations of the theory can be obtained by taking a variation of Eq.~\eqref{actiongeo} with respect to the metric $g_{MN}$, which yields
\begin{eqnarray}
\frac{\partial f}{\partial R}R_{MN}&-&\frac{1}{2}f\left(R,T\right)g_{MN}-\left(\nabla_M\nabla_N-g_{MN}\Box\right)\frac{\partial f}{\partial R}
	\nonumber \\
&=&\kappa^2 T_{MN}-\frac{\partial f}{\partial T}\left(T_{MN}+\Theta_{MN}\right),\label{fieldgeo}
\end{eqnarray}
where $\nabla_M$ denotes covariant derivatives and $\Box\equiv\nabla^M\nabla_M$ is the d'Alembert operator, both written in terms of the metric $g_{MN}$. The stress-energy tensor $T_{MN}$ is defined in terms of the variation of the matter Lagrangian density $\mathcal L_m$ in the usual way, i.e., 
\begin{equation}\label{defTab}
T_{MN}=-\frac{2}{\sqrt{-g}}\frac{\delta\left(\sqrt{-g}\mathcal L_m\right)}{\delta g^{MN}},
\end{equation}
and $\Theta_{MN}$ is a tensor defined in terms of the variation of the stress-energy tensor $T_{MN}$ with respect to the metric $g_{MN}$ as
\begin{equation}\label{deftheta}
\Theta_{MN}=g^{PQ}\frac{\delta T_{PQ}}{\delta g^{MN}}.
\end{equation}
The explicit form of the tensor $\Theta_{MN}$ can only be obtained after the form of the stress-energy tensor $T_{MN}$ is defined (or, equivalently, the form of the matter Lagrangian $\mathcal L_m$). 

A dynamically equivalent scalar-tensor representation of the action in Eq. \eqref{actiongeo} can be obtained via the introduction of two auxiliary fields $\alpha$ and $\beta$ as
\begin{eqnarray}
S=\frac{1}{2\kappa^2}\int_\Omega\sqrt{-g}\left[f\left(\alpha,\beta\right)+\frac{\partial f}{\partial \alpha}\left(R-\alpha\right)\right.
	\nonumber\\
\left.+\frac{\partial f}{\partial\beta}\left(T-\beta\right)\right]d^5x+S_m\left(g_{MN},\chi\right).\label{auxaction1}
\end{eqnarray}
The action \eqref{auxaction1} depends on three independent variables, namely, the metric $g_{MN}$ and the two auxiliary fields $\alpha$ and $\beta$. Taking the variations of Eq.~\eqref{auxaction1} with respect to $\alpha$ and $\beta$ yields the two equations of motion
\begin{equation}\label{auxeom1}
f_{\alpha\alpha}\left(R-\alpha\right)+f_{\alpha\beta}\left(T-\beta\right)=0,
\end{equation}
\begin{equation}\label{auxeom2}
f_{\beta\alpha}\left(R-\alpha\right)+f_{\beta\beta}\left(T-\beta\right)=0,
\end{equation}
respectively, where the subscripts $\alpha$ and $\beta$ denote partial derivatives with respect to these fields and $f_{\alpha\beta}=f_{\beta\alpha}$ as we assume the function $f\left(\alpha,\beta\right)$ is well-behaved and thus satisfies the Schwartz theorem. The system of Eqs. \eqref{auxeom1} and \eqref{auxeom2} can be recast in a matricial form $\mathcal M \textbf{X}=0$ as
\begin{equation}\label{matrixeq}
\mathcal M \textbf{X}=\begin{pmatrix}
f_{\alpha\alpha} & f_{\alpha\beta} \\
f_{\beta\alpha} & f_{\beta\beta} 
\end{pmatrix}
\begin{pmatrix}
R-\alpha\\
T-\beta
\end{pmatrix}
=0.
\end{equation}
The solution of the system of Eq. \eqref{matrixeq} will be unique if and only if the determinant of the matrix $\mathcal M$ is non-vanishing. The condition $\text{det}\mathcal M\neq 0$ yields a constraint between the second-order derivatives of $f$ in the form $f_{\alpha\alpha}f_{\beta\beta}\neq f_{\alpha\beta}^2$. If this condition is satisfied, the solution for the system of Eqs.~\eqref{auxeom1} and \eqref{auxeom2} is unique and given by $\alpha=R$ and $\beta=T$. One can now verify that inserting these considerations back into Eq.~\eqref{auxaction1}, one recovers the original action in Eq.~\eqref{actiongeo}, thus confirming the consistency of this transformation and proving the equivalence between the two representations.

If one now defines two scalar fields $\varphi$ and $\psi$ and a scalar interaction potential $V\left(\varphi,\psi\right)$ in the forms
\begin{equation}\label{defscalar}
\varphi=\frac{\partial f}{\partial R}, \qquad \psi=\frac{\partial f}{\partial T},
\end{equation}
\begin{equation}\label{defpotential}
V\left(\varphi,\psi\right)=\varphi R+\psi T-f\left(R,T\right),
\end{equation}
the auxiliary action \eqref{auxaction1} can be rewritten in the equivalent scalar-tensor representation as
\begin{eqnarray}
S=\frac{1}{2\kappa^2}\int_\Omega\sqrt{-g}\left[\varphi R+\psi T-V\left(\varphi,\psi\right)\right]d^5x\nonumber \\
+S_m\left(g_{MN},\chi\right).\label{actionst}
\end{eqnarray}
The action in Eq. \eqref{actionst} depends on three independent variables, namely, the metric $g_{MN}$ and the two scalar fields $\varphi$ and $\psi$. Taking the variation of Eq.~\eqref{actionst} with respect to these variables, one obtains respectively
\begin{eqnarray}
&&\varphi R_{MN}-\frac{1}{2} g_{MN}\left(\varphi R+\psi T-V\right)
-\left(\nabla_M\nabla_N-g_{MN}\Box\right)\varphi
 \nonumber \\
&&\qquad \qquad \qquad =\kappa^2 T_{MN}-\psi\left(T_{MN}+\Theta_{MN}\right),
\label{fieldst}
\end{eqnarray}
\begin{equation}\label{eomphi}
V_\varphi=R,
\end{equation}
\begin{equation}\label{eompsi}
V_\psi=T,
\end{equation}
where the subscripts $\varphi$ and $\psi$ denote partial derivatives with respect to these fields, respectively. Notice that Eq.~\eqref{fieldst} could be obtained directly from Eq.~\eqref{fieldgeo} by introducing directly the definitions in Eqs.~\eqref{defscalar} and \eqref{defpotential}.

We remark that, in the formalism described by Eqs.~~\eqref{defscalar}--\eqref{eompsi}, one can see that the standard case, $f(R,T)=R$ is recovered for $\varphi=1$ and $\psi=0$, giving $V(\varphi,\psi)=0$. In this situation, there is no variation with respect to the fields $\varphi$ and $\psi$, so Eqs.~\eqref{eomphi} and \eqref{eompsi} do not exist, and the problem is described by Eq.~\eqref{fieldst}, which becomes
$R_{MN}-g_{MN} R/2= \kappa^2 T_{MN}$.

\subsection{Matter distribution}
Let us now consider matter to be described by a single dynamical scalar field $\chi$ with an interaction potential $U\left(\chi\right)$. The matter action that describes this distribution of matter is given by
\begin{equation}\label{actionchi}
S_m=-\int_\Omega\sqrt{-g}\left[\frac{1}{2}\partial^P\chi\partial_P\chi+U\left(\chi\right)\right]d^5x.
\end{equation}
Taking a variation of Eq.~\eqref{actionchi} with respect to the scalar field $\chi$ and using the definition of the stress-energy tensor $T_{MN}$ provided in Eq.~\eqref{defTab} yields
\begin{equation}\label{tmn}
T_{MN}=-g_{MN}\left[\frac{1}{2}\partial^P\chi\partial_P\chi+U\left(\chi\right)\right]+\partial_M\chi\partial_N\chi.
\end{equation}
The explicit form of $T_{MN}$ obtained in Eq.~\eqref{tmn} allows one to finally compute the associated form of $\Theta_{MN}$ via the definition in Eq.~\eqref{deftheta}. Taking the variation of Eq.~\eqref{tmn} with respect to the inverse metric $g^{MN}$ one obtains
\begin{equation}\label{theta}
\Theta_{MN}=g_{MN}\left[\frac{1}{2}\partial^P\chi\partial_P\chi+U\left(\chi\right)\right]-\frac{5}{2}\partial_M\chi\partial_N\chi.
\end{equation}

Furthermore, an equation of motion for the field $\chi$ can now be obtained by taking a variation of Eq.~\eqref{actionst} with respect to $\chi$. This equation of motion will not only feature terms arising from the matter action $S_m$ but also terms arising from the gravitational sector, as it depends explicitly on the trace of $T_{MN}$ in Eq.~\eqref{tmn}. The resultant equation of motion takes the form 
\begin{equation}\label{eomchi}
\left(\frac{3\psi}{2\kappa^2}+1\right)\Box\chi+\frac{3}{2\kappa^2}\partial_a\chi\partial^a\psi=\left(\frac{5\psi}{2\kappa^2}+1\right)U_\chi,
\end{equation}
where the subscript $\chi$ denotes a derivative with respect to the scalar field $\chi$.

\subsection{Metric and equations of motion}

To investigate branes, we consider that the spacetime manifold is described by the 5-dimensional line element
\begin{equation}\label{metric}
ds^2=e^{2A\left(y\right)}\eta_{\mu\nu}dx^\mu dx^\nu+dy^2,
\end{equation}
where $A\left(y\right)$ is called the warp function, $\eta_{\mu\nu}=\text{diag}\left(-1,1,1,1\right)$ is the 4-dimensional Minkowski metric, where greek indeces run from $0$ to $3$, and $y$ represents an extra infinite 5th dimension. Furthermore, let us assume that the gravitational scalar fields $\varphi$, $\psi$ and the matter scalar field $\chi$ are constant through the 4-dimensional spacetime and depend solely on the extra dimension $y$, i.e., $\varphi=\varphi\left(y\right)$, $\psi=\psi\left(y\right)$, and $\chi=\chi\left(y\right)$. We see that the three scalar fields $\varphi$, $\psi$, and $\chi$ act as source fields to generate brane configurations in the present braneworld scenario, also guided by the warp function $A$ and the potentials $V$ and $U$. Given the isotropy of the 4-dimensional part of the metric, Eq.~\eqref{fieldst} will yield only two independent field equations. Inserting  Eqs.~\eqref{metric} into Eqs.~\eqref{fieldst}--\eqref{eompsi} and \eqref{eomchi}, along with the forms of $T_{MN}$ and $\Theta_{MN}$ from Eqs.~\eqref{tmn} and \eqref{theta} one obtains the system
\begin{eqnarray}\label{field1}
&&3\varphi\left(2A'^2+A''\right)+3\varphi'A'+\frac{V}{2}+\varphi''
	\nonumber \\
&&\qquad =-\left(\kappa^2+\frac{5\psi}{2}\right)U-\left(\kappa^2+\frac{3\psi}{2}\right)\frac{\chi'^2}{2},
\end{eqnarray}
\begin{eqnarray}\label{field2}
6\varphi A'^2+4\varphi' A'+\frac{V}{2}
&=&-\left(\kappa^2+\frac{5\psi}{2}\right)U
	\nonumber\\
&&+\left(\kappa^2+\frac{3\psi}{2}\right)\frac{\chi'^2}{2},
\end{eqnarray}
\begin{equation}\label{eqphi}
V_\varphi=-20A'^2-8A'',
\end{equation}
\begin{equation}\label{eqpsi}
V_\psi=-5U-\frac{3}{2}\chi'^2,
\end{equation}
\begin{equation}\label{eqchi}
\left(\frac{3\psi}{2\kappa^2}+1\right)\left(4A'\chi'+\chi''\right)+\frac{3}{2\kappa^2}\chi'\psi'=\left(\frac{5\psi}{2\kappa^2}+1\right)U_\chi ,
\end{equation}
where the prime denotes a derivative with respect to the coordinate $y$. It is important to note that in the system of Eqs.~\eqref{field1}--\eqref{eqchi} only four of these five equations are independent. This feature can be demonstrated by taking the derivative of Eq.~\eqref{field2} with respect to $y$, using Eq.~\eqref{field1} to cancel the term dependent on $A''$, using Eq.~\eqref{eqphi} to cancel $V_\varphi$, using Eq.~\eqref{eqpsi} to cancel $V_\psi$, and finally using Eq.~\eqref{field2} itself to cancel the terms dependent on $A'$, thus obtaining Eq.~\eqref{eqchi} as a result. 

To recover the standard case easily, one can substitute Eq.~\eqref{field1} by the difference between Eqs.~\eqref{field1} and \eqref{field2}, in the form
\begin{equation}\label{field1alt}
3\varphi A''-\varphi'A'+\varphi''= -\left(\kappa^2+\frac{3\psi}{2} \right){\chi'}^2,
\end{equation}
and use this equation instead of Eq.~\eqref{field1} to investigate the problem. As we commented below Eq.~\eqref{eompsi}, the equations for the standard case, $f(R,T)=R$, are obtained with $\varphi=1$ and $\psi=0$, and $V(\varphi,\psi)=0$. Thus, Eqs.~\eqref{eqphi} and \eqref{eqpsi} do not exist and the above equation becomes 
\begin{equation}
A''= -\frac13\kappa^2{\chi'}^2.
\end{equation}
Also, the Eq.~\eqref{field2} takes the form 
\begin{equation}
{A'}^2=-\frac16\kappa^2 U + \frac1{12}\kappa^2 {\chi'}^2,
\end{equation}
and Eq.~\eqref{eqchi} simplifies to 
\begin{equation}
\chi''+ 4A'\chi'= U_\chi,
\end{equation}
which is the equation for the field $\chi$. As it is known, only two of these three equations are independent. In this situation, considering $\kappa^2=2$ as usual, by introducing an auxiliary function, $W(\chi)$, associated to the potential which is now defined by
\begin{equation}
U(\chi) = \frac12W_\chi^2-\frac{4}{3}W^2,
\end{equation}
one can obtain first order equations, in the form 
\begin{equation}
\chi'= W_\chi \qquad A'=-\frac{2}{3}W.
\end{equation}
In $f(R,T)$ gravity, it was shown in Ref.~\cite{Bazeia:2015owa} that first order equations can be obtained in models in which $f(R,T)=R+\gamma T$. It is interesting to see that, in the action of Eq.~\eqref{actionst} this model is obtained with $\varphi=1$, $\psi=\gamma$, and $V=0$.

\subsection{Brane stability}\label{sec:stab}

The study of small perturbations of the metric was previously done in Ref.~\cite{Bazeia:2015owa} in the usual representation, given by $f(R,T)$ in the action in Eq. \eqref{actiongeo}. There, the authors have shown that by taking
\begin{equation}\label{metricpert}
ds^2=e^{2A\left(y\right)}\left[\eta_{\mu\nu}+H_{\mu\nu}(x,y)\right] dx^\mu dx^\nu + dy^2,
\end{equation}
where $H_{\mu\nu}(x,y)$ is a small perturbation around a Minkowski background metric $\eta_{\mu\nu}$, the gravity sector of the brane is linearly stable for $f(R,T) = F(R) + G(T)$. In the scalar-tensor representation described by Eqs.~~\eqref{defscalar}--\eqref{eompsi}, this situation is equivalent to the case $V(\varphi,\psi)= P(\varphi) + Q(\psi)$. By following Ref.~\cite{Bazeia:2015owa}, one gets the stability equation
\begin{equation}\label{stab}
\left[-\frac{d^2}{dz^2} + u(z) \right]\bar H_{\mu\nu}(z) = p^2 \bar H_{\mu\nu}(z),
\end{equation}
where the variable $z$ was defined as $dz=e^{-A(y)}dy$ to make the metric conformally flat, and one considers the transverse traceless components of metric fluctuation $H_{\mu\nu}$, which is written in terms of $\bar H_{\mu\nu}$ as $H_{\mu\nu}(x,z)=e^{-ip\cdot x}e^{-3A(z)/2 }\varphi^{-1/2}\bar H_{\mu\nu}(z)$. In the above equation, the stability potential has the form
\begin{equation}\label{stabpot}
u\left(z\right)=\alpha\left(z\right)^2-\frac{d\alpha}{dz},
\end{equation}
where the function $\alpha\left(z\right)$ is defined as
\begin{equation}
\alpha\left(z\right)=-\frac{3}{2}A_z-\frac{1}{2}\frac{d}{dz}\left(\ln\varphi\right).
\end{equation}
The Schr\"odinger-like equation in Eq.~\eqref{stab} can be factorized in the form $S^\dagger S\bar H_{\mu\nu} = p^2 \bar H_{\mu\nu}$, where
\begin{equation}
S=\frac{d}{dz}+\alpha\left(z\right)\qquad S^\dagger=-\frac{d}{dz}+\alpha\left(z\right).
\end{equation}
The aforementioned factorization ensures that $p^2\geq0$, so the gravity sector of the brane is stable.

The above expression allows us to calculate the massless graviton state, represented by the zero mode, $p^2 = 0$. By taking $S \bar H_{\mu\nu}^{(0)}=0$, we get
\begin{equation}\label{zeromode}
\bar H_{\mu\nu}^{(0)} = N_{\mu\nu}\sqrt{\varphi\left(z\right)}e^{3A(z)/2},
\end{equation}
where $N_{\mu\nu}$ is a normalization factor. Notice that, in the standard case, $\varphi=1$ and the zero mode is ${\tilde{H}}_{\mu\nu}^{(0)}=\tilde{N}_{\mu\nu} e^{3A(z)/2}$. The normalization factors $N_{\mu\nu}$ can be obtained via the integration of the zero mode as
\begin{equation}\label{zerointegral}
\int \bar H_{\mu\nu}^{(0)}dz=N_{\mu\nu}\int \varphi e^{2A}dy=1,
\end{equation}
which guarantees that the 4-dimensional gravity can be recovered on the brane.

\section{Solutions in the general case}\label{sec:solsgen}

Let us now focus attention on the presence of solutions in some specific cases of general interest and on the stability of the gravitational sector. We consider, in particular, the simpler case with $\chi=0$, and another one, in which the matter field $\chi$ also plays a role. 

\subsection{Solution without matter ($\chi=0$)}\label{sec:model1}

We start our analysis with the simplest possible case of a brane model without the matter field $\chi$, i.e., solely supported by the gravitational scalar fields $\varphi$ and $\psi$. In this case, one assumes $\chi=0$ and $U=0$ for the matter scalar field and potential. Under these assumptions, Eq.~\eqref{eqchi} is automatically satisfied. Furthermore, from Eq.~\eqref{eqpsi} one verifies that $V_\psi=0$ and thus the potential $V$ must be solely a function of $\varphi$, i.e., $V\left(\varphi,\psi\right)=V\left(\varphi\right)$. This result allows one to use the chain rule to write $V_\varphi$ as a function of $V'\left(y\right)$ and $\varphi'$. Under these considerations, Eq. \eqref{field1alt} and the equation of motion for $\varphi$ in Eq.~\eqref{eqphi} become respectively
\begin{equation}\label{m1field}
3\varphi A''+\varphi''-A'\varphi'=0,
\end{equation} 
\begin{equation}\label{m1phi}
A''+\frac{5}{2}A'^2+\frac{1}{8}\frac{V'}{\varphi'}=0.
\end{equation}

Equations \eqref{m1field} and \eqref{m1phi} are a system of two independent equations for the three unknowns $A$, $\varphi$ and $V$. Thus, the system is under-determined and one can still impose one constraint to close the system. Since we are interested in thick-brane solutions, we chose to set an explicit form for the warp function $A$ as
\begin{equation}
A\left(y\right)=A_0 \log\left[\text{sech}\left(ky\right)\right],
\end{equation}
where $A_0$ and $k$ are constant parameters and $A_0$ in particular must be positive-defined. The system of Eqs.~\eqref{m1field} and \eqref{m1phi} thus take the following form
\begin{equation}\label{m1field2}
\varphi''+k A_0\tanh\left(ky\right)\varphi'-3k^2A_0\ \text{sech}^2\left(ky\right)\varphi=0,
\end{equation}
\begin{equation}\label{m1phi2}
\frac{V'}{\varphi'}=4k^2A_0\left[2 \text{sech}^2\left(ky\right)-5A_0\tanh^2\left(ky\right)\right],
\end{equation}
respectively.

\begin{figure*}[htb!]
\includegraphics[scale=0.95]{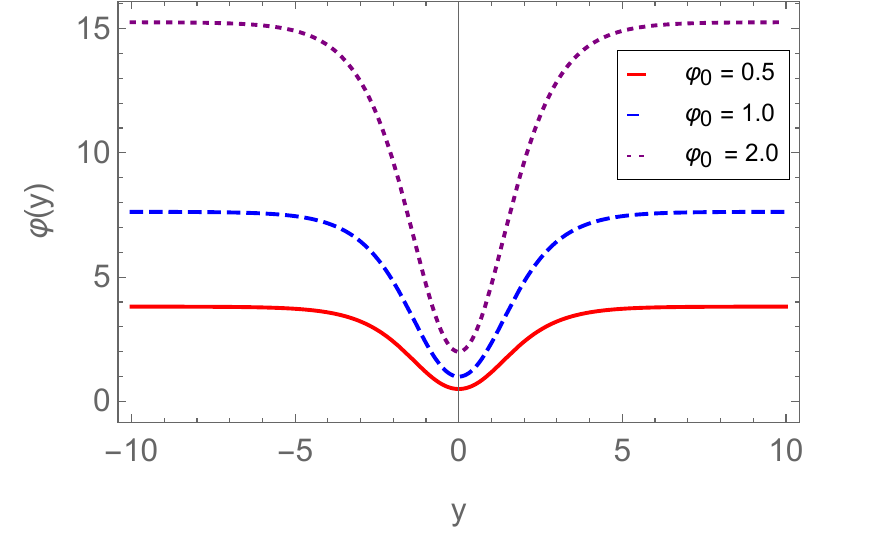}
\includegraphics[scale=0.95]{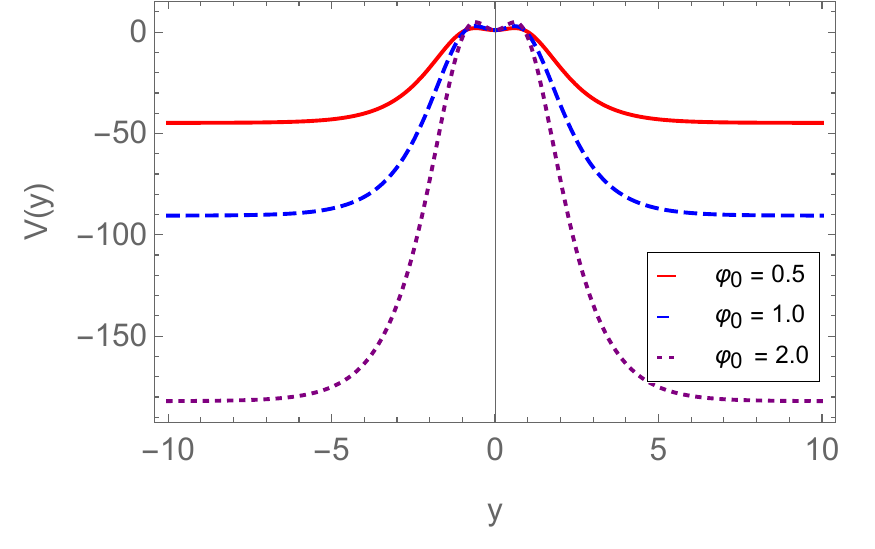}
\caption{Solutions for $\varphi\left(y\right)$ (left plot) and $V\left(y\right)$ (right plot) resulting from the integration of Eqs.~\eqref{m1field2} and \eqref{m1phi2} with $A_0=1$, $k=1$, and $V_0=1$, for different values of $\varphi_0$.}
\label{fig:model1}
\end{figure*}

Equations \eqref{m1field2} and \eqref{m1phi2} form a system of two coupled differential equations for $\varphi$ and $V$. Due to their complexity, these equations do not feature analytical solutions and must be solved numerically. In particular, one starts by solving Eq.~\eqref{m1field2} for $\varphi$ and then inserts the result into Eq.~\eqref{m1phi2} to solve to $V$. These solutions must satisfy a set of boundary conditions at the origin that guarantee that they are even, i.e., $\varphi\left(0\right)=\varphi_0$, $V\left(0\right)=V_0$, where $\varphi_0$ and $V_0$ are constants, and $\varphi'\left(0\right)=0$ and $V'\left(0\right)=0$. The numerical solutions for this case are plotted in Fig.~\ref{fig:model1}, where we have considered $A_0=1$, $k=1$ and $V_0=1$ for simplicity. The solutions for $\varphi$ grow outwards from $y=0$ attaining a constant asymptotic value as $y\to \pm \infty$. The solutions for $V$ start growing outwards from $y=0$ forming a small potential well but eventually reverse their growth and decrease outwards to attain a lower asymptotic value at $y\to \pm\infty$.

\begin{figure*}[htb!]
\includegraphics[scale=0.9]{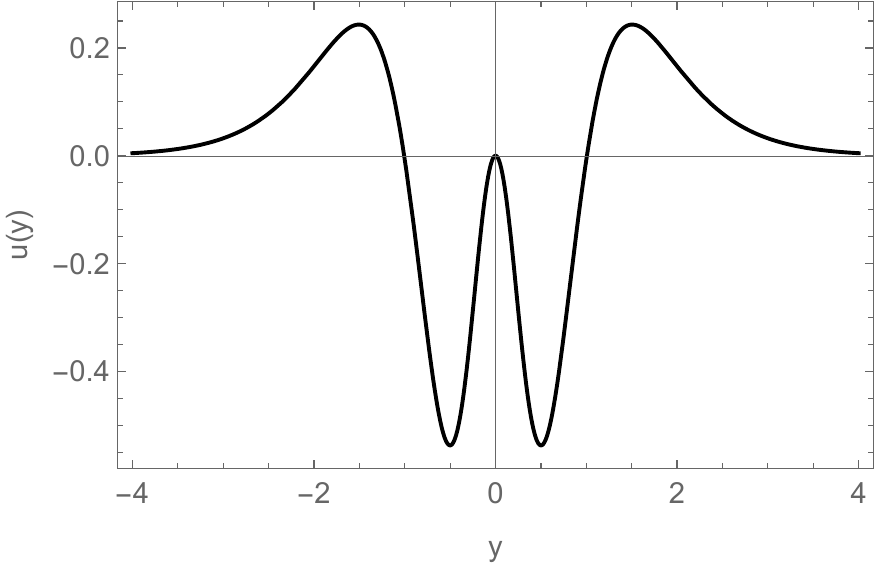}
\ \ \ \ \ 
\includegraphics[scale=0.9]{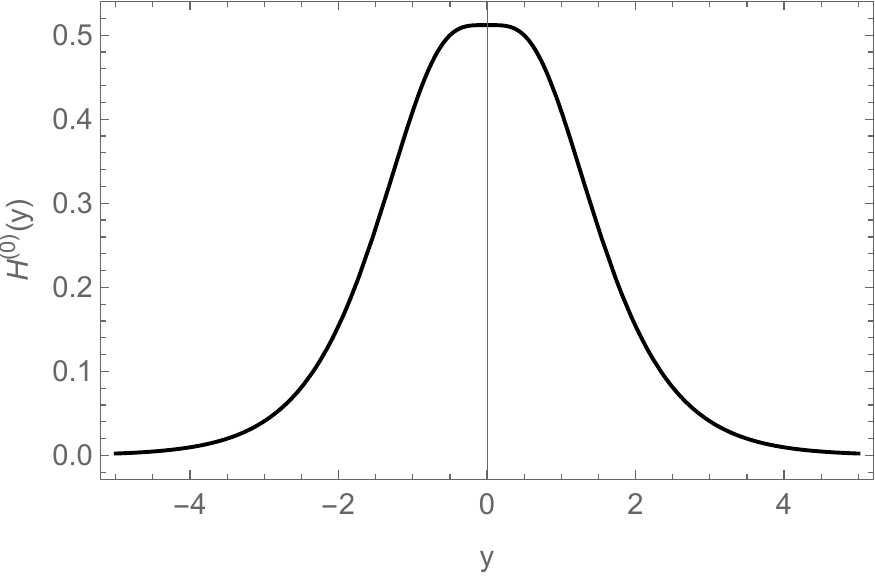}
\caption{Stability potential $u\left(y\right)$ (left panel) and graviton zero mode $H_{\mu\nu}^{(0)}$ (right panel) for the model proposed in Sec.~\ref{sec:model1} for $k=1$, $A_0=1$, $\varphi_0=1$, and $V_0=1$. These results remain unchanged for different values of $\varphi_0$ and $V_0$.}
\label{fig:modes1}
\end{figure*}

For this model without matter, since $f(R,T)$ depends only on $R$, we can see from Sec.~\ref{sec:stab} that the gravity sector of the brane is stable. Therefore, we can use Eqs.~~\eqref{stabpot} and \eqref{zeromode} to calculate the stability potential $u$ and the graviton zero mode $H_{\mu\nu}^{(0)}$, which are plotted in Fig.~\ref{fig:modes1}. The potential $u$ vanishes at $y=0$, decreases outwards achieving two global minima at some $|y|=y_{\text{min}}$, and then proceeds to increase attaining two global maxima at some $|y|=y_{\text{max}}$, with $y_{\text{min}}<y_{\text{max}}$, this presenting a double potential well. The general behavior of the potential $u\left(y\right)$ is not affected by different choices of the free parameters: changes in $\varphi_0$ and $V_0$ do not induce any modifications in the potential, and changes in $A_0$ and $k$ produce simple rescalings of the potential without changing its general shape. Since the potential vanishes at $y=0$, the graviton zero mode $H_{\mu\nu}^{(0)}$ presents a single peak on the brane, and thus the brane does not develop an internal structure.

\subsection{Solution with matter ($\chi\neq 0$)}\label{sec:model2}

Let us now consider solutions for thick branes in the presence of a matter scalar field $\chi$. Equations \eqref{field2}--\eqref{field1alt} form a system of five equations, of which only four are independent, to the six unknowns $A$, $\varphi$, $\psi$, $V$, $\chi$, and $U$. Furthermore, as will be shown later in this section, since in this case there are no restrictions in the form of $V$, i.e., it will be a general function of two variables $\varphi$ and $\psi$, there are in fact two degrees of freedom associated with this unknown, and thus one has effectively a total of seven degrees of freedom hidden in the six unknowns. This means that the system is under-determined and one can impose three constraints to close the system. Note however that setting an explicit form of the potential would reduce the number of degrees of freedom by two, and consequently only one extra constraint could be imposed. In this case, we will impose explicit forms of the matter field $\chi$ and the potential $U$ as
\begin{equation}\label{m2chi}
\chi\left(y\right)=\chi_0 \tanh\left(ky\right),
\end{equation}
\begin{equation}\label{m2U}
U\left(\chi\right)=\frac{1}{2}W_\chi^2-\frac{4}{3}W\left(\chi\right)^2,
\end{equation}
where $\chi_0$ and $k$ are arbitrary constants and the function $W\left(\chi\right)$ is called the super-potential of $\chi$ and is given by the form
\begin{equation}\label{m2W}
W\left(\chi\right)=\chi-\frac{1}{3}\chi^3.
\end{equation}
The motivation behind the choices made in Eqs.~\eqref{m2chi} and \eqref{m2U} resides in their wide use in the literature due to their close connection to the standard case in General Relativity. On the other hand, as the model in study features a non-vanishing Ricci scalar $R$ and trace of the stress-energy tensor $T$, from Eqs.~\eqref{eqphi} and \eqref{eqpsi} one verifies that $V_\varphi$ and $V_\psi$ are non-zero, and thus the potential $V$ will depend explicitly in both scalar fields $\varphi$ and $\psi$. As a consequence, the derivative of $V$ with respect to $y$ will be given by the chain rule
\begin{equation}\label{poteq}
V'\left(y\right)=V_\varphi\left(y\right)\varphi'+V_\psi\left(y\right)\psi'.
\end{equation}
This result shows that there is a degeneracy associated to the potential $V\left(y\right)$: one can use Eq.~\eqref{eqphi} to obtain $V_\varphi\left(y\right)$ as a function of $A$, then use Eq.~\eqref{eqpsi} to obtain $V_\psi\left(y\right)$ as a function of $y$, and replace these results into Eq.~\eqref{poteq}. One is thus left with a system of three independent equations, the equation of motion for $\chi$ in Eq.~\eqref{eqchi}, Eq.~\eqref{field1alt}, and the potential equation just derived in Eq.~\eqref{poteq} for the four unknowns $A$, $V$, $\varphi$ and $\psi$. 

The origin of this degeneracy is the fact that there are two degrees of freedom associated with the potential V, even though it is a single unknown. This fact can already be inferred from Eq. \eqref{poteq}, which shows that one can always consider either $V_\varphi$ or $V_\psi$ to be independent from $V$ as there are infinite different combinations of $V_\varphi$ and $V_\psi$ that yield the same $V$. Let us nevertheless prove this degeneracy explicitly: consider the system of Eqs. \eqref{field1} to \eqref{eqchi} plus Eq. \eqref{poteq}, and take $V_\varphi$ and $V_\psi$ as functions of $y$. Taking the derivative of Eq. \eqref{field2}, using Eq. \eqref{eqchi} to eliminate $\chi$, using Eq. \eqref{field1} to eliminate $\varphi''$, using Eq. \eqref{eqphi} to eliminate $A''$, using Eq. \eqref{eqchi} to eliminate $U$, using Eq. \eqref{field2} again to eliminate $V$, and finally using Eq. \eqref{poteq} to eliminate $V'$, one obtains an identity. This proves that one of the equations in the system of Eqs. \eqref{field1} to \eqref{eqchi} along with Eq. \eqref{poteq} is not independent, and thus they form a system of five equations for the eight unknown functions of $y$: $A$, $\varphi$, $\psi$, $\chi$, $U$, $V$, $V_\varphi$, and $V_\psi$. If at this point one leaves $V$, $V_\varphi$ and $V_\psi$ arbitrary, one is able to introduce three constraints in the remaining variables to close the system. On the other hand, if one imposes an explicit form of $V\left(\varphi,\psi\right)$, this also sets the forms of $V_\varphi$ and $V_\psi$, and Eq. \eqref{poteq} is identically satisfied, thus leaving us with a system of four equations to five unknowns, and allowing only for second constraint. Thus, as we have given explicit forms of $\chi$ and $U$, this degeneracy in the potential effectively allow us to introduce one extra constraint to close the system, as long as this constraint is not the form of $V$. Similarly to the case without matter, we chose to impose an explicit form of the warp factor $A$ as
\begin{equation}
A\left(y\right)=A_0 \log\left[\text{sech}\left(ky\right)\right],
\end{equation}
where $A_0$ is a positive-defined constant. The resultant closed system of equations can then be written as the subtraction of Eq.~\eqref{field2} from Eq.~\eqref{field1}, the equation of motion for $\chi$ in Eq.~\eqref{eqchi}, and the potential equation in Eq.~\eqref{poteq}, which take the respective forms
\begin{eqnarray}
\varphi'' &=& -\frac{1}{2}k\left[+2A_0\tanh\left(ky\right)\varphi'-6kA_0 \text{sech}^2\left(ky\right)\varphi\right.
	\nonumber \\
&&+\left. k\chi_0^2 \text{sech}^4\left(ky\right)\left(4+3\psi\right)\right],\label{m2field}
\end{eqnarray}
\begin{eqnarray}
&&\frac{3}{4}k\chi_0 \text{sech}^2\left(ky\right)\psi'+\frac{1}{18}\left\{84-100\chi_0^2\tanh^2\left(ky\right)\right. 
	\nonumber \\
&&+\left.16\chi_0^4\tanh^4\left(ky\right)-k^2\left(36+27\psi\right)\left(1+2A_0\right) \text{sech}^2\left(ky\right)\right.
	\nonumber\\
&&+\left.\left[84-25\chi_0^2\tanh^2\left(ky\right)+4\chi_0^4\tanh^4\left(ky\right)\right]\psi\right\}=0,\label{m2chi2}
\end{eqnarray}
\begin{eqnarray}
V'&=&2k^2A_0\left[4+5A_0-5A_0\cosh\left(2ky\right)\right] \text{sech}^2\left(ky\right)\varphi'
	\nonumber\\
&&-\frac{3}{2}k^2\chi_0^2 \text{sech}^4\left(ky\right)+\frac{5}{54}\left[-27+126\chi_0^2\tanh^2\left(ky\right) \right. 
	\nonumber\\
 &&-\left.75\chi_0^4\tanh^4\left(ky\right)+8\chi_0^6\tanh^6\left(ky\right)\right].\label{m2pot}
\end{eqnarray}

\begin{figure*}[htb!]
\includegraphics[scale=0.95]{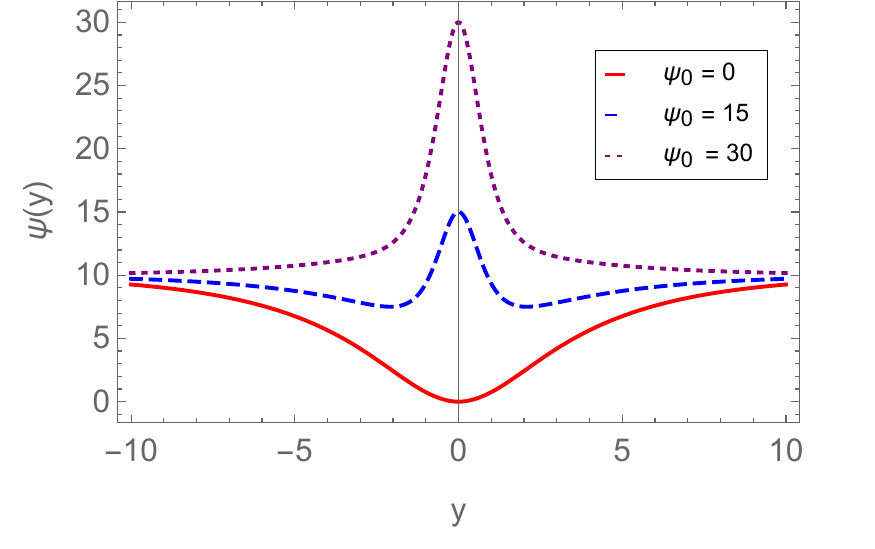}
\includegraphics[scale=0.95]{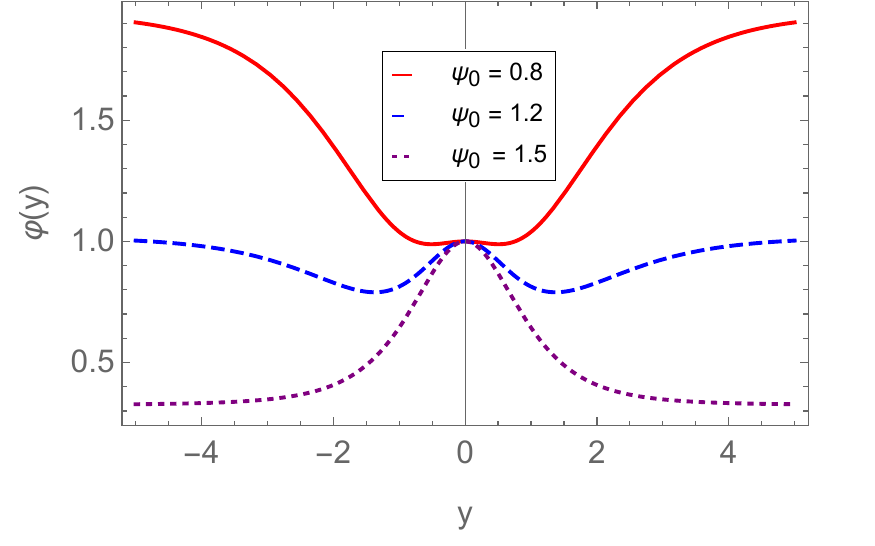}
\caption{Solutions for $\psi\left(y\right)$ (left panel) and $\varphi\left(y\right)$ (right panel) resulting from the integration of Eqs.~\eqref{m2field} to \eqref{m2pot} with $A_0=1$, $k=1$, and $V_0=1$. In these plots we also consider $\varphi_0=1$ for simplicity, but the same plethora of behaviors could be obtained for any other value of $\varphi_0$.}
\label{fig:model2}
\end{figure*}

\begin{figure}[htb!]
\includegraphics[scale=0.95]{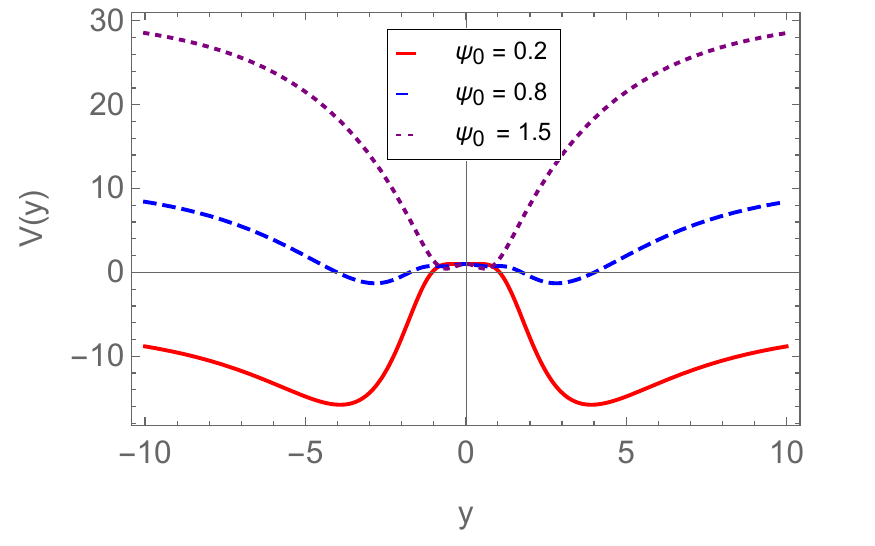}
\caption{Solutions for $V\left(y\right)$ resulting from the integration of Eqs.~\eqref{m2field} to \eqref{m2pot} with $A_0=1$, $k=1$, and $V_0=1$. In these plots we also consider $\varphi_0=1$ for simplicity, but the same plethora of behaviors could be obtained for any other value of $\varphi_0$.}
\label{fig:model2b}
\end{figure}

Equations \eqref{m2field}--\eqref{m2pot} constitute a system of three coupled differential equations for $\varphi$, $\psi$ and $V$ that must be solved numerically subjected to appropriate boundary conditions at the origin that guarantee the evenness of the solutions, i.e., $\varphi\left(0\right)=\varphi_0$, $\psi\left(0\right)=\psi_0$, and $V\left(0\right)=V_0$, where $\varphi_0$, $\psi_0$ and $V_0$ are constants, and also $\varphi'\left(0\right)=0$, $\psi'\left(0\right)=0$, and $V'\left(0\right)=0$. The numerical solutions for this model are plotted in Figs.~\ref{fig:model2} and \ref{fig:model2b}, where we have again considered $A_0=1$, $k=1$, and $V_0=0$ for the same reasons as outlined in the previous model. The scalar field $\psi$ can present at most three different behaviors depending solely in the value of $\psi_0$: it may increase outwards from the origin (e.g. for $\psi_0=0$), it may decrease outwards from the origin, attain a minimum value at some $|y|=y_\text{min}$, and proceed to increase again all the way up to $y\to\pm\infty$ (e.g. for $\psi_0=15$, or it may have a global maximum at $y=0$ and decrease monotonically outwards from the origin (e.g. for $\psi_0=30$). 

\begin{figure}[htb!]
\includegraphics[scale=0.9]{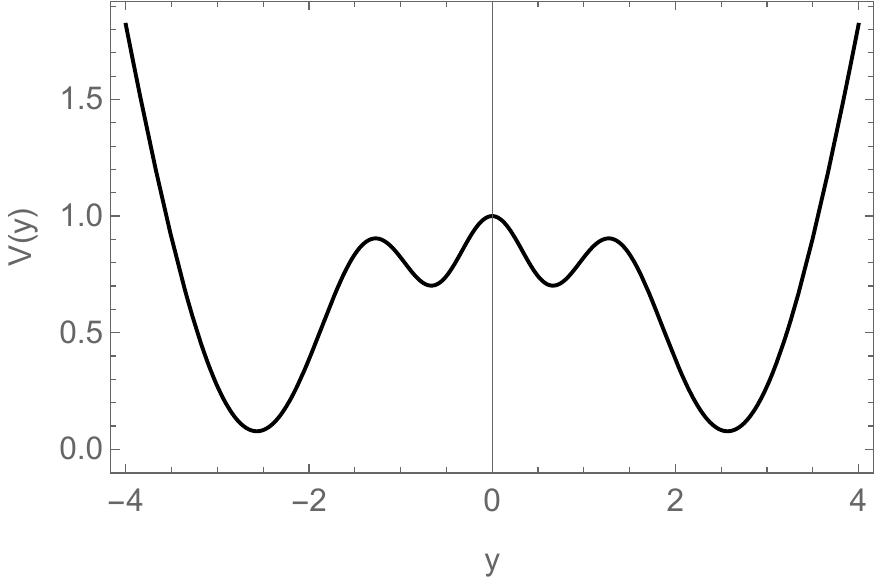}
\caption{Solution for $V\left(y\right)$ from the integration of Eq.~\eqref{m2pot} with $A_0=1$, $k=1$, $V_0=1$, $\varphi_0=1$, and $\psi_0=0.87$. The four-minima structure that the potential $V$ acquires for $\varphi_0=1$ and $0.7\lesssim\psi_0\lesssim0.95$ is visible.}
\label{fig:potspec}
\end{figure}

It was also verified that for a fixed value of $\varphi_0$ there are ranges of values of $\psi_0$ for which the behaviors of the scalar field $\varphi$ and the potential $V$ change dramatically. Consider as an example the case $\varphi_0=1$. For this case, if $\psi_0\lesssim 0.8$ the scalar field $\varphi$ is monotonically decreasing outwards from the origin, if $0.8\lesssim \psi_0 \lesssim 1.5$ the scalar field $\varphi$ decreases outwards from the origin, attains two minima at some $|y|=y_\text{min}$, and increases again outwards, and if $\psi_0\gtrsim 1.5$ the scalar field $\psi$ has a global  minimum at the origin and increases monotonically outwards. Also for $\varphi_0=1$, one verifies that if $\psi_0\lesssim 0.7$ the potential $V$ decreases outwards from the origin, attains two minima at some $|y|=y_\text{min}$, and proceeds to grow outwards. At $\psi_0\sim 0.7$ the potential $V$ develops two new minima at some $|y|=\bar y_\text{min}$ with $\bar y_\text{min} < y_\text{min}$. This structure with four minima is maintained for the range $0.7\lesssim \psi_0\lesssim 0.95$, see Fig.~\ref{fig:potspec} for an explicit example. At $\psi_0\sim 0.95$, the minima at $|y|=y_\text{min}$ collapse into saddle points, and for $\psi_0 \gtrsim 0.95$ the potential again recovers a behavior with only two minima at the points $|y|=\bar y_\text{min}$.

\begin{figure*}[htb!]
\includegraphics[scale=0.96]{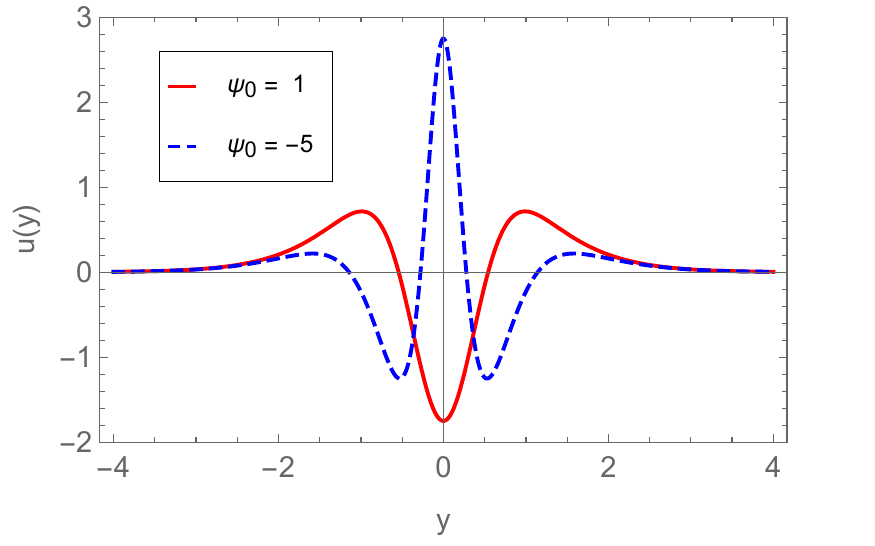}
\includegraphics[scale=0.95]{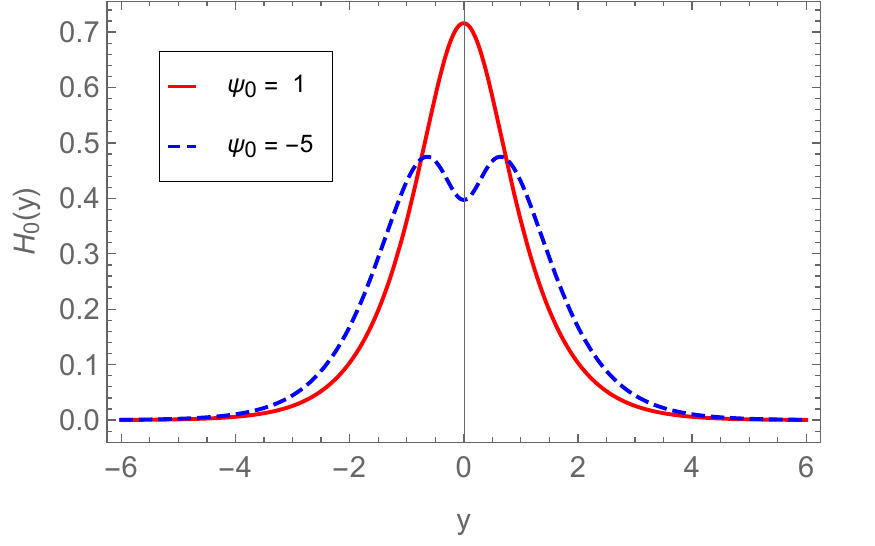}
\caption{Stability potential $u\left(y\right)$ (top panel) and graviton zero mode $H_{\mu\nu}^{(0)}$ (bottom panel) for the model proposed in Sec.~\ref{sec:model2} for $k=1$, $A_0=1$, $\varphi_0=1$, and $V_0=1$, for different values of $\psi_0$.}
\label{fig:modes2}
\end{figure*}

We have shown that the model investigated in this section supports brane solutions. However, $V$ was obtained as a function of $y$ and, since there are two fields $\varphi$ and $\psi$ involved in the model, it is not possible to determine the explicit form of $V$ as a function of $\varphi$ and $\psi$. Indeed, because of the degeneracy of the potential $V$, the form $V\left(\varphi,\psi\right)$ is not unique. To draw conclusions about the stability of the model, it is necessary that the potential $V$ is separable in terms of $\varphi$ and $\psi$ (see Sec.~\ref{sec:stab}). Thus, one can exploit the degeneracy of the potential supposing that $V(\varphi,\psi) = P(\varphi) + Q(\psi)$. In this case, 
$V'\left(y\right)=(dP/d\varphi)\left(y\right)\varphi'+ (dQ/d\psi)\left(y\right) \psi',$ and Eq. \eqref{poteq} implies that $V_\varphi\left(y\right)=(dP/d\varphi)\left(y\right)$ and $V_\psi\left(y\right)= (dQ/d\psi)\left(y\right)$, which can be done without loss of generality as $V_\varphi$, $V_\psi$, $dP/d\varphi$ and $dQ/d\psi$ are functions of $y$. In other words, any solution in the general case $f\left(R,T\right)$ obtained with an arbitrary $V\left(\varphi,\psi\right)$ can be recast as a solution of a separable potential, which implies that the gravity sector of the brane is stable, as discussed in Sec.~\ref{sec:stab}. We then use Eqs.~\eqref{stabpot} and \eqref{zeromode} to calculate the stability potential $u$ and the graviton zero mode $H_{\mu\nu}^{(0)}$, which are plotted in Fig.~\ref{fig:modes2} for different combinations of the free parameters. Again, one verifies that the parameter $V_0$ does not induce any changes in the solutions and that a modification on the parameters $k$ and $A_0$ simply re-scales the solutions without altering their general behavior. In this case, however, the parameters $\varphi_0$ and $\psi_0$ play a crucial role in defining the shape of the potential $u$ and consequently the shape of the zero mode $H_{\mu\nu}^{(0)}$: for a given value of $\varphi_0>0$ set to guarantee the positiveness of the term proportional to $R$ in the action of Eq.~\eqref{actionst}, one can tune the value of $\psi_0$ to control whether the potential $u$ presents a single potential well, a double potential well, or even a potential barrier in the brane. Consider e.g. $k=1$, $A_0=1$, $V_0=1$ and $\chi_0=1$, and set $\varphi_0=1$. For this combination of parameters, one verifies that if $\psi_0\gtrsim -0.6$ the potential $u$ presents a single potential well on the brane and the graviton zero mode $H_{\mu\nu}^{(0)}$ presents a single peak on the brane; when $-1.3\lesssim\psi_0\lesssim-0.6$ the potential $u$ develops a negative local minimum on the brane which flattens the peak of the zero mode $H_{\mu\nu}^{(0)}$ on the brane; and finally if $\psi_0\lesssim -1.3$ the potential $u$ develops a potential barrier on the brane surrounded by two potential wells at some $|y|=y_{\text{min}}$ which effectively breaks the graviton zero mode $H_{\mu\nu}^{(0)}$ in two separate peaks, case for which the brane does support an internal structure.

\section{Other possibilites}
\label{sec:other}

We now investigate other models, in which the function $f\left(R,T\right)$ is restricted to be linear in either $R$ or $T$, and the stability of their gravity sector is ensured from the beginning. These models must be analyzed separately from the general case because they do not satisfy the condition $\det\mathcal M\neq 0$, for the matrix $\mathcal M$ defined in Eq.~\eqref{matrixeq}, and thus the scalar-tensor formalism has to be changed. In the following we pursue the analysis of these cases.

\subsection{The case $f\left(R,T\right)=F\left(R\right)+T$}\label{sec:theocase1}

Let us now consider a special case for which the function $f\left(R,T\right)$ can be decomposed in the form $F\left(R\right)+T$, for some arbitrary function $F\left(R\right)$. The second order partial derivatives of this function are $f_{RR}=F_{RR}$, $f_{TT}=f_{RT}=0$, and thus we have $f_{RR}f_{TT}-f_{RT}^2=0$. Therefore, the general method described in Eqs.~\eqref{auxaction1}--\eqref{eompsi} is not well-defined and we have to analyze this case independently. The action in Eq.~\eqref{actiongeo} takes the form
\begin{equation}\label{actioncase1}
S=\frac{1}{2\kappa^2}\int_\Omega\sqrt{-g}\left[F\left(R\right)+T\right]d^5x+S_m\left(g_{MN},\chi\right).
\end{equation}
To obtain a dynamically equivalent scalar-tensor representation for this action we only need one auxiliary field $\alpha$, since the degree of freedom associated with the arbitrary dependency of $f\left(R,T\right)$ in $T$ is no longer present. We can thus write the geometrical part of Eq.~\eqref{actioncase1}, i.e, ignoring the matter action for simplicity as it does not play any role in the transformation that follows, in the form
\begin{equation}\label{auxactcase1}
S=\frac{1}{2\kappa^2}\int_\Omega\sqrt{-g}\left[F\left(\alpha\right)+T+\frac{dF}{d\alpha}\left(R-\alpha\right)\right]d^5x.
\end{equation}
The action in Eq.~\eqref{auxactcase1} now depends on two independent variables, namely the metric $g_{MN}$ and the auxiliary field $\alpha$. Taking a variation with respect to $\alpha$ yields the equation of motion
\begin{equation}\label{eomalpha}
F_{\alpha\alpha}\left(R-\alpha\right)=0.
\end{equation}
This result implies that the solution of Eq.~\eqref{eomalpha} is unique only if the function $F\left(\alpha\right)$ is at least quadratic in $\alpha$. In that case, the unique solution becomes $R=\alpha$ and Eq.~\eqref{auxactcase1} reduces to Eq.~\eqref{actioncase1}, thus proving the equivalence of the two representations. One can now define a scalar field $\varphi$ and a potential $V\left(\varphi\right)$ in the forms
\begin{equation}
\varphi=\frac{dF}{dR}, \qquad V\left(\varphi\right)=\varphi R-F(R),
\end{equation}
and obtain the equivalent scalar-tensor representation of the $f\left(R,T\right)$ gravity in the particular case $f\left(R,T\right)=F\left(R\right)+T$ as
\begin{equation}\label{actionstcase1}
S=\frac{1}{2\kappa^2}\int_\Omega\sqrt{-g}\left[\varphi R+ T-V\left(\varphi\right)\right]d^5x+S_m\left(g_{MN},\chi\right).
\end{equation}
Equation \eqref{actionstcase1} now depends on two independent variables, the metric $g_{MN}$ and the scalar field $\varphi$, and one can derive two equations of motion. Taking the variation of Eq.~\eqref{actionstcase1} with respect to $g_{MN}$ and $\varphi$ respectively yields
\begin{eqnarray}
&&\varphi R_{MN}-\frac{1}{2} g_{MN}\left(\varphi R+ T-V\right)
 \label{fieldstcase1} \\
&&-\left(\nabla_M\nabla_N-g_{MN}\Box\right)\varphi=\kappa^2 T_{MN}-\left(T_{MN}+\Theta_{MN}\right),\nonumber
\end{eqnarray}
\begin{equation}\label{eomphicase1}
V_\varphi=R.
\end{equation}
Notice that these equations could be obtained directly from Eqs.~\eqref{fieldst} and \eqref{eomphi} by taking $\psi=1$ and $V\left(\varphi,\psi\right)=V\left(\varphi\right)$, but this situation can not be obtained as a limit of the general case because the remaining equation, i.e., Eq.~\eqref{eompsi}, would force $T=0$, whereas in here the matter distribution remains arbitrary.

Let us now look for solutions in the current case, with the matter distribution in Eq.~\eqref{actionchi} and the metric in Eq. \eqref{metric}. From Eqs.~\eqref{fieldstcase1}, \eqref{eomphicase1} and \eqref{eomchi} with $\psi=1$, we get the following independent equations 
\begin{equation}\label{fieldcase1}
3\varphi A''-\varphi'A'+\varphi''=-\left(\kappa^2+\frac{3}{2}\right)\chi'^2,
\end{equation}
\begin{equation}\label{kgphicase1}
V'\left(y\right)=-(20A'^2+8A'')\varphi',
\end{equation}
\begin{equation}\label{eomchicase1}
\left(\frac{3}{2\kappa^2}+1\right)\left(4A'\chi'+\chi''\right)=\left(\frac{5}{2\kappa^2}+1\right)U_\chi,
\end{equation}
where Eq.~\eqref{fieldcase1} arrives from the subtraction of the $(y,y)$ component from the $(t,t)$ component of the field equations in Eq.~\eqref{fieldstcase1} and we have used the chain rule to write $V_\varphi(\varphi)$ in terms of $V'\left(y\right)$ and $\varphi'$.

\begin{figure}[htb!]
\includegraphics[scale=0.9]{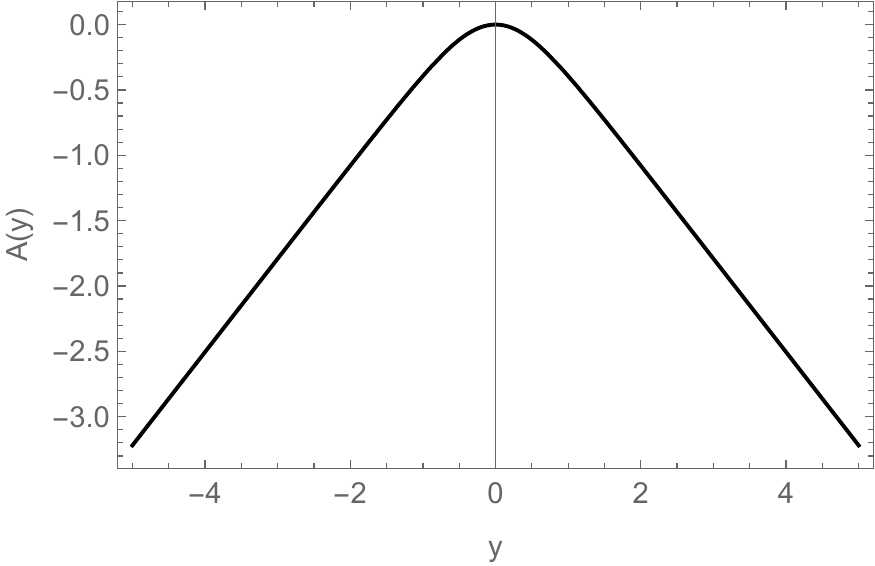}
\caption{Solution for $A\left(y\right)$ obtained from the integration of Eqs.~\eqref{fieldcase1}--\eqref{eomchicase1} under the assumptions of Eqs.~\eqref{m2chi}--\eqref{m2W} with $\chi_0=1$ and $k=1$.}
\label{fig:zeta3}
\end{figure}

The system of Eqs.~\eqref{fieldcase1} to \eqref{eomchicase1} is a system of three independent equations to the five unknowns $\varphi$, $A$, $V$, $U$ and $\chi$, which implies that one can impose two constraints to determine the system. We chose to set the forms of $\chi$ and $U$ as in Eqs.~\eqref{m2chi}--\eqref{m2W}. Under these considerations, Eq.~\eqref{eomchicase1} decouples from the rest and can be directly integrated to find a solution for $A$ with an arbitrary integration constant $A_0$ which is set according to the boundary condition $A\left(0\right)=0$. The solution for $A\left(y\right)$ is analytical but, given its lengthy expression and for the sake of clarity, we decided to plot this solution in Fig.~\ref{fig:zeta3} instead of writing its explicit form. Afterwards, this solution for $A$ is introduced into Eq.~\eqref{fieldcase1} which can then be solved for $\varphi$ subject to the boundary conditions $\varphi\left(0\right)=\varphi_0$ and $\varphi'\left(0\right)=0$ for some arbitrary constant $\varphi_0$. Finally, one can introduce the solutions for $A$ and $\varphi$ into Eq.~\eqref{kgphicase1} and solve for $V$ considering the boundary condition $V\left(0\right)=V_0$. 

\begin{figure*}[htb!]
\includegraphics[scale=0.95]{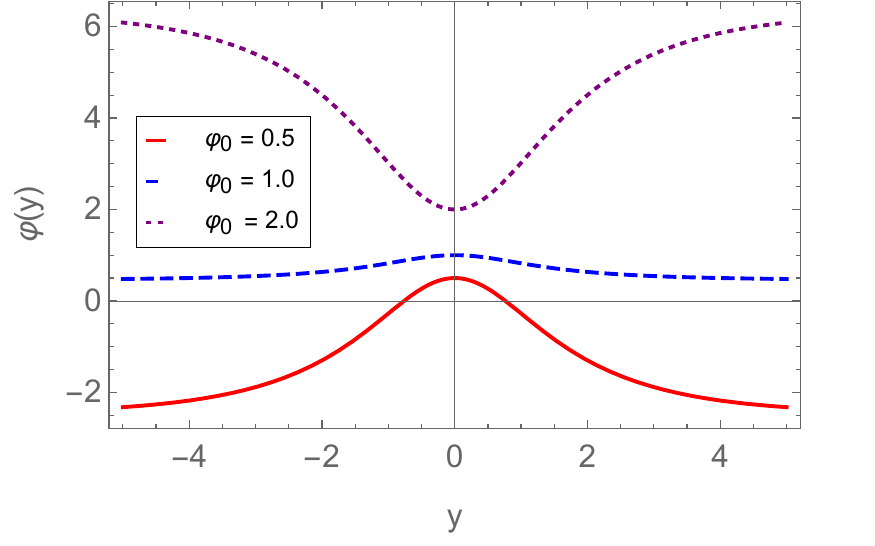}
\includegraphics[scale=0.95]{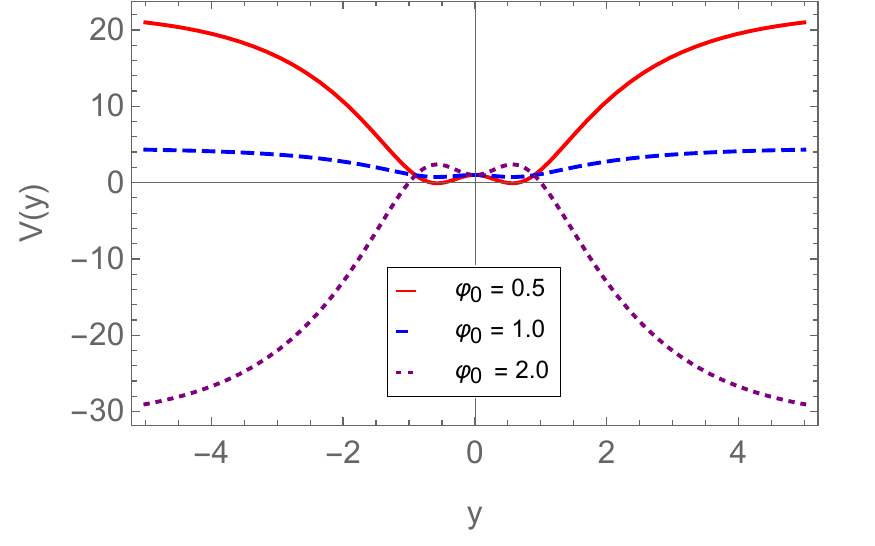}
\caption{Solutions for $\varphi\left(y\right)$ (left panel) and $V\left(y\right)$ (right panel) resulting from the integration of Eqs.~\eqref{fieldcase1} and \eqref{kgphicase1} with $\chi_0=1$, $k=1$, and $V_0=1$. It can be seen that the behaviors of $\varphi$ and $V$ are strongly correlated.}
\label{fig:model3}
\end{figure*}

The numerical solutions for $\varphi$ and $V$ in this model are plotted in Fig.~\ref{fig:model3} where we have considered $\chi_0=1$, $k=1$, and $V_0=1$, as it was verified that these parameters do not influence the general behavior of the solutions, functioning solely as scaling factors. It was verified that the shape of the potential $V$ varies with the choice of $\varphi_0$, but the shape of $V_0$ does not affect the shape of $\varphi$, as expected since the solutions for $\varphi$ was obtained from an equation that decouples completely from $V$. The solution for for $\varphi$ can either attain a global maximum at $y=0$ and decrease outwards, which happens for $\varphi_0\lesssim 1$, or attain a global minimum at $y=0$ and increase outwards, which happens for $\varphi_0\gtrsim 1$. Accordingly, the shape of the potential $V$ can either be a double barrier with a small potential well at $y=0$ or a double well with a small potential barrier at $y=0$, respectively. Notice how in this case, since the potential depends solely in one scalar field, the behaviors of the two quantities $\varphi$ and $V$ are strongly connected.

\begin{figure*}[htb!]
\includegraphics[scale=0.95]{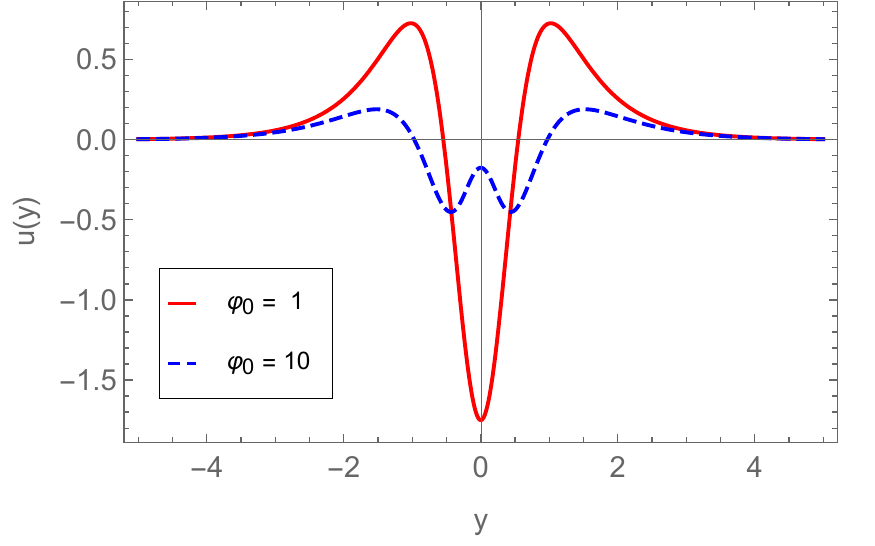}
\ \ \ \ \ 
\includegraphics[scale=0.95]{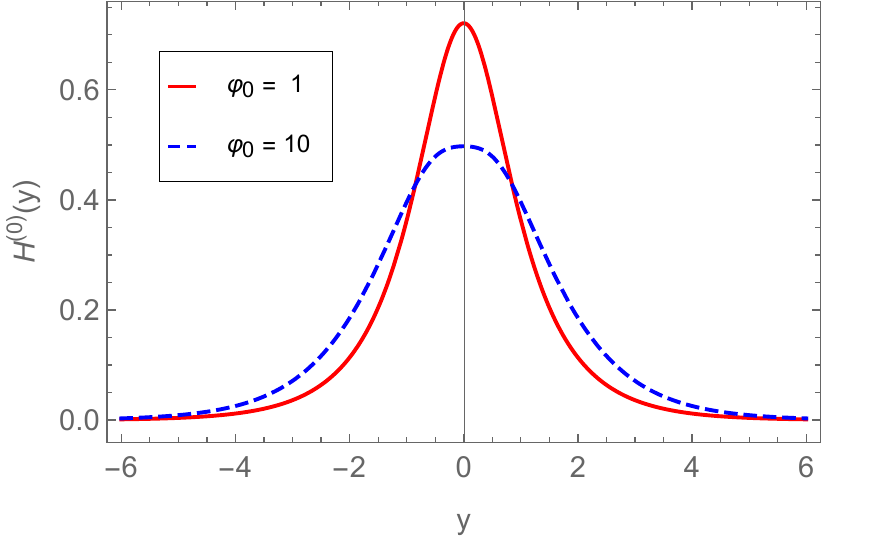}
\caption{Stability potential $u\left(y\right)$ (left panel) and graviton zero mode $H_{\mu\nu}^{(0)}$ (right panel) for the model proposed in Sec.~\ref{sec:theocase1} for $k=1$, $\chi_0=1$, and $V_0=1$, for different values of $\varphi_0$.}
\label{fig:modes3}
\end{figure*}

Since $V$ depends only on $\varphi$, the model supports brane configurations with stable gravity sector, as suggested by the discussion in Sec.~\ref{sec:stab}. This feature allows us to calculate the solutions for the stability potential $u$ in Eq.~\eqref{stabpot} and the gravity zero mode $H_{\mu\nu}^{(0)}$ in Eq.~\eqref{zeromode}, which are plotted in Fig.~\ref{fig:modes3}. It was verified that the parameter $V_0$ does not affect the results and that $\chi_0$ and $k$ function solely as rescaling factors, and thus these parameters were fixed at $V_0=1$, $\chi_0=1$, and $k=1$. For this choice of parameters, one verifies that for small values of $\varphi_0$ in the range $\varphi_0\lesssim 3.5$ the potential $u$ presents a global minimum at $y=0$, increases outwards attaining two global maxima at some $|y|=y_{\text{max}}$, and proceeds to decrease outwards again. In the regime $\varphi\gtrsim 3.5$, a local maximum develops at $y=0$ separating the global minimum into two potential wells at some $|y|=y_{\text{min}}$, with $y_{\text{min}}<y_{\text{max}}$. This local maximum increases with $\varphi_0$ but tends to zero as $\varphi_0\to\infty$, thus never developing a potential barrier. Consequently, the gravity zero mode $H_{\mu\nu}^{(0)}$ is always characterized by a single central maximum at $y=0$, and the brane does not develop internal structure.

\subsection{The case $f\left(R,T\right)=R+G\left(T\right)$}\label{sec:theocase2}

Consider now another particular case of interest for which the function $f\left(R,T\right)$ is decomposed in the form $R+G\left(T\right)$, for a given function $G\left(T\right)$. The second order partial derivatives of this function are $f_{RR}=f_{RT}=0$, $f_{TT}=G_{TT}$, and thus again we have $f_{RR}f_{TT}-f_{RT}^2=0$. The general method described before in Eqs.~\eqref{auxaction1}--\eqref{eompsi} is not well-defined and again we have to analyze this case separately. The action in Eq.~\eqref{actiongeo} takes the form
\begin{equation}\label{actioncase2}
S=\frac{1}{2\kappa^2}\int_\Omega\sqrt{-g}\left[R+G\left(T\right)\right]d^5x+S_m\left(g_{MN},\chi\right).
\end{equation}
Similarly to the previous particular case, a dynamically equivalent scalar-tensor representation for this action requires only one auxiliary field $\beta$, as the degree of freedom associated with the arbitrary dependency of $f\left(R,T\right)$ in $R$ ceases to exist. We can thus write the geometrical part of Eq.~\eqref{actioncase2}, i.e, again ignoring the matter action as it is not relevant in the transformation that follows, in the form
\begin{equation}\label{auxactcase2}
S=\frac{1}{2\kappa^2}\int_\Omega\sqrt{-g}\left[R+G\left(\beta\right)+\frac{dG}{d\beta}\left(T-\beta\right)\right]d^5x.
\end{equation}
The action in Eq.~\eqref{auxactcase2} now depends on two independent variables, namely the metric $g_{MN}$ and the auxiliary field $\beta$. Performing a variation with respect to $\beta$ leads to the equation of motion
\begin{equation}\label{eombeta}
G_{\beta\beta}\left(T-\beta\right)=0.
\end{equation}
Thus, the solution of Eq.~\eqref{eomalpha} is unique only if the function $G\left(\beta\right)$ is at least quadratic in $\beta$. If this condition is verified, the unique solution becomes $T=\beta$ and Eq.~\eqref{auxactcase2} reduces to Eq.~\eqref{actioncase2}, thus proving the equivalence of the two representations. We can now define a scalar field $\psi$ and a potential $V\left(\psi\right)$ in the forms
\begin{equation}
\psi=\frac{dG}{dT}, \qquad V\left(\psi\right)= \psi T-G\left(T\right),
\end{equation}
and obtain the equivalent scalar-tensor representation of the $f\left(R,T\right)$ gravity in the particular case in which  $f\left(R,T\right)=R+G\left(T\right)$ as
\begin{equation}\label{actionstcase2}
S=\frac{1}{2\kappa^2}\int_\Omega\sqrt{-g}\left[R+ \psi T-V\left(\psi\right)\right]d^5x+S_m\left(g_{MN},\chi\right).
\end{equation}
Equation \eqref{actionstcase2} now depends on two independent variables, the metric $g_{MN}$ and the scalar field $\psi$, and again we can obtain two equations of motion. Taking the variation of Eq.~\eqref{actionstcase2} with respect to $g_{MN}$ and $\psi$ gives, respectively,
\begin{equation}\label{fieldstcase2}
R_{MN}-\frac{1}{2} g_{MN}\left(R\!+\!\psi T\!-\!V\right)=\kappa^2 T_{MN}-\psi\left(T_{MN}\!+\!\Theta_{MN}\right),
\end{equation}
and
\begin{equation}\label{eompsicase2}
V_\psi=T.
\end{equation}
Again, we note that these equations could also be obtained directly from Eqs.~\eqref{fieldst} and \eqref{eompsi} by taking $\varphi=1$ and $V\left(\varphi,\psi\right)=V\left(\psi\right)$, but one can not obtain this situation as a limit of the general case because the remaining equation, i.e., Eq.~\eqref{eomphi}, would force the solutions to have $R=0$, whereas in here the geometry remains arbitrary.

To investigate brane configurations, we consider the matter distribution in Eq.~\eqref{actionchi} and the metric from Eq. \eqref{metric}. From Eqs.~\eqref{fieldstcase2}, \eqref{eompsicase2} and \eqref{eomchi}, we get the following independent equations
\begin{equation}\label{fieldcase2}
3A''= -\left(\kappa^2+\frac{3\psi}{2} \right){\chi'}^2.
\end{equation}
\begin{equation}\label{kgpsicase2}
V'\left(y\right)=-\left(5U+\frac{3}{2}\chi'^2\right)\psi',
\end{equation}
\begin{equation}\label{eomchicase2}
\left(\frac{3\psi}{2\kappa^2}+1\right)\left(4A'\chi'+\chi''\right)+\frac{3}{2\kappa^2}\chi'\psi'=\left(\frac{5\psi}{2\kappa^2}+1\right)U_\chi,
\end{equation}
where Eq.~\eqref{fieldcase2} is again obtained from the difference between the $(y,y)$ and the $(t,t)$ components of the field equations in Eq.~\eqref{fieldstcase2} and we have used the chain rule to write $V'\left(\psi\right)$ in terms of $V'\left(y\right)$ and $\psi'$. The system of Eqs.~\eqref{fieldcase2} to \eqref{eomchicase2} is a system of three independent equations for five unknowns $\psi$, $A$, $V$, $U$ and $\chi$, which implies that two constraints can be imposed to close the system. Similarly to what we did before, we set the forms of $\chi$ and $U$  as in Eqs. \eqref{m2chi}--\eqref{m2U}. These ansatze do not decouple the system of Eqs.~\eqref{fieldcase2}--\eqref{eomchicase2} and thus the solutions must be obtained via simultaneous numerical integrations under appropriate choices of boundary conditions. The property $A\left(0\right)=0$ is imposed as a boundary condition for $A$, and we also impose $A'\left(0\right)=0$ to preserve the parity of the solutions. Furthermore, we impose $\psi\left(0\right)=\psi_0$ and $V\left(0\right)=V_0$ for some arbitrary constants $\psi_0$ and $V_0$.

\begin{figure*}[htb!]
\includegraphics[scale=0.95]{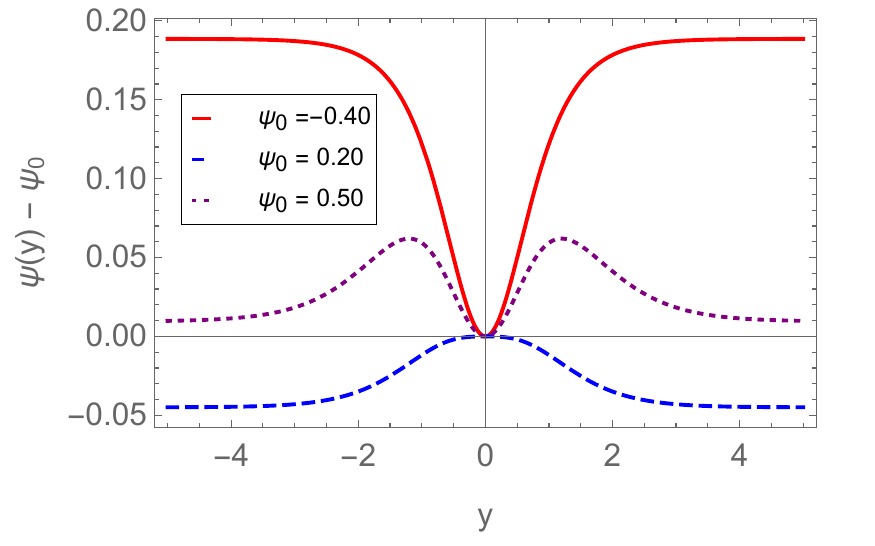}
\includegraphics[scale=0.95]{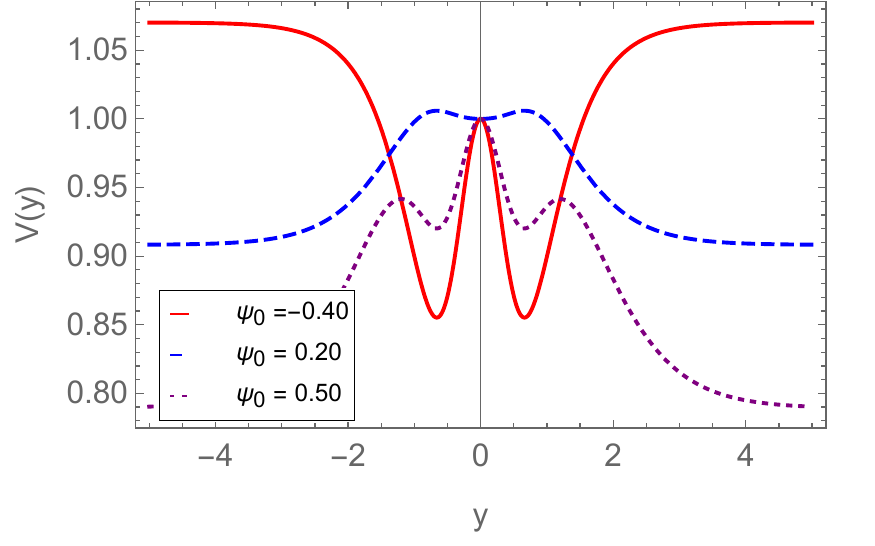}
\caption{Solutions for $\psi\left(y\right)$ (left panel) and $V\left(y\right)$ (right panel) resulting from the integration of Eqs.~\eqref{fieldcase2}--\eqref{eomchicase2} with $\chi_0=1$, $k=1$, and $V_0=1$. It can be seen that the behaviors of $\psi$ and $V$ are strongly correlated.}
\label{fig:model4}
\end{figure*}

The numerical solutions for $\psi$ and $V$ are plotted in Fig.\ref{fig:model4} where we have considered $\chi_0=1$, $k=1$, and $V_0=1$, as these parameters do not influence the general behavior of the solutions. In this case, we chose not to plot explicitly the solution for $A$ since the behavior is nearly identical to the previous case already plotted in Fig.\ref{fig:zeta3}. Furthermore, since $\psi_0$ functions simply as a rescaling factor for $A$, we have plotted $A\left(y\right)$ for a single value $\psi_0=1$. Again, we verified that the value of $V_0$ does not affect the solutions for $\psi_0$, as the equations from which the solution for $\psi_0$ is computed is independent of $V$. On the other hand, the behavior of $V$ can present numerous different properties and it is strongly dependent on the choice of $\psi_0$. The solutions for $\psi$ and $V$ can have at most three different behaviors: in the regime $\psi_0<0$, $\psi$ attains a global minimum at $y=0$ and increases outwards, whereas $V$ presents a double-well structure with a potential barrier at $y=0$; for $0<\psi_0\lesssim 0.26$ both $\psi$ and $V$ invert their behaviors, i.e., $\psi$ attains a global maximum at $y=0$ and decreases outwards, whereas $V$ adopts a double barrier structure with a small potential well at $y=0$; and when $\psi_0\gtrsim 0.26$, $\psi$ develops a minimum at $y=0$, increases outwards to attain two global maxima at some $|y|=y_{\text{max}}$, and proceeds to decrease outwards, while the potential $V$ develops a complex structure with a triple potential barrier. Again, it is remarkable that, since the potential becomes a function of a single scalar field in this particular case, the behaviors of the two quantities $\psi$ and $V$ become highly correlated.

\begin{figure*}[htb!]
\includegraphics[scale=0.95]{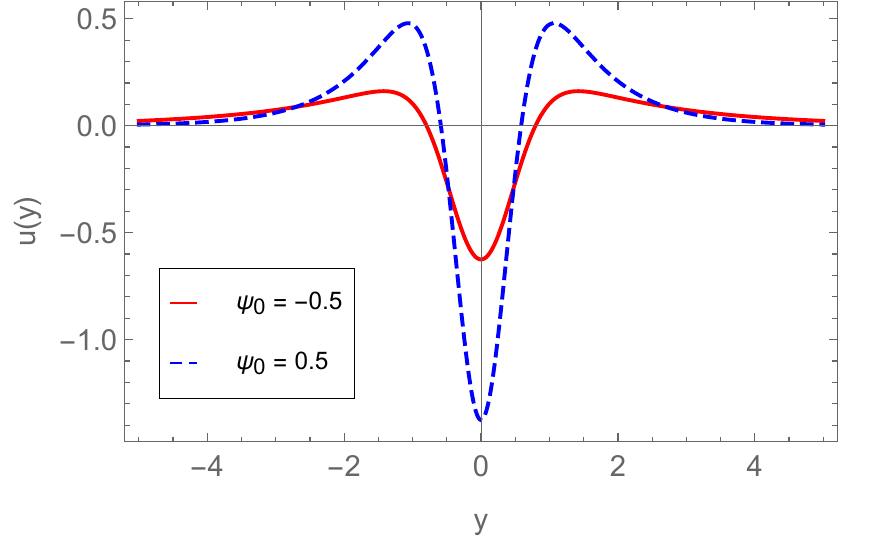}
\ \ \ \ \ 
\includegraphics[scale=0.95]{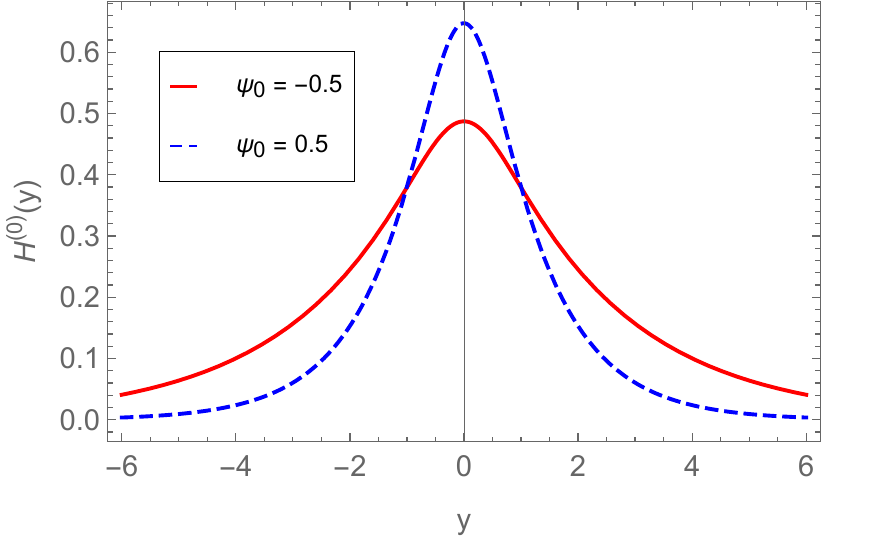}
\caption{Stability potential $u\left(y\right)$ (left panel) and graviton zero mode $H_{\mu\nu}^{(0)}$ (right panel) for the model proposed in Sec.~\ref{sec:theocase2} for $k=1$, $\chi_0=1$, and $V_0=1$, for different values of $\psi_0$.}
\label{fig:modes4}
\end{figure*}

Let us analyze the stability of this model. Since $V$ depends only on $\psi$, we can ensure the stability of the gravity sector of the brane. We then use Eqs.~\eqref{stabpot} and \eqref{zeromode} to calculate the stability potential $u$ and the graviton zero mode $H_{\mu\nu}^{(0)}$ numerically and plot them in Fig.~\ref{fig:modes4}. Since this model is characterized by a constant scalar field $\varphi=1$, Eqs.~\eqref{stabpot} and \eqref{zeromode} imply that the stability potential and the zero mode are controlled uniquely by the warp function $A$, which was shown in Fig.~\ref{fig:zeta3} to have the usual behavior. Thus, even though there is a clear dependency of the stability potential and the zero mode in the parameter $\psi_0$, in this case this parameter effectively becomes a simple rescaling factor, similarly to the parameters $\chi_0$ and $k$. Furthermore, the solutions remain unaffected by the parameter $V_0$. As a consequence, $u$ only presents a single well behavior on the brane $y=0$ and the zero mode $H_{\mu\nu}^{(0)}$ is characterized by a single central peak.

\section{Conclusions}\label{sec:concl}

In this work, we have studied the presence of braneworld solutions in modified theories of gravity in a 5D bulk in the presence of an extra spatial dimension of infinite extent. We have considered the possibility to construct solutions in the scalar-tensor formulation of gravity, modified by the presence of functions of the Ricci scalar $R$ and the trace of the stress-energy tensor $T$, in the form $f(R,T)$. We have first developed the general procedure to investigate the problem, which requires the presence of two real scalar fields, $\varphi$ and $\psi$, and then studied some specific situations, such as in the presence of an extra scalar field. 

To study the stability of the gravity sector of the models, we have considered the previous result introduced in \cite{Bazeia:2015owa}, in which the stability is proven when the model is written in the case where $f(R,T)=F(R)+G(T)$. We have then taken advantage of this result and considered two distinct new possibilities, with $f(R,T)= F(R)+T$ and with $f(R,T)=R+G(T)$. In this sense, we adapted the general formalism, which requires the presence of the two fields $\varphi$ and $\psi$, to the case in the presence of a single field, $\varphi$ or $\psi$. In these specific cases the gravitational sectors of the brane are also stable, and in the models studied we have also displayed the corresponding stability potential and the gravity zero mode.

The results achieved in Secs. \ref{sec:solsgen} and \ref{sec:other} are of particular interest. In Sec. \ref{sec:solsgen} we have shown that for the general case of a function $f\left(R,T\right)$ featuring two scalar degrees of freedom it is possible to obtain solutions for braneworlds presenting an internal structure. On the other hand, if one chooses particular forms of the function $f\left(R,T\right)$ for which only one of the scalar degrees of freedom is present, like it was studied in Sec. \ref{sec:other}, these braneworld solutions with internal structure are unattainable. This result seems to confirm that internal structure is a characteristic sourced by a system of two scalar fields, and none is singly responsible for its development, which is consistent with the results previously published for the Bloch brane and the hybrid metric-Palatini gravity $f\left(R,\mathcal R\right)$. An interesting line of investigation that opens up with the presence of an internal structure is related to the possibility of making the extra dimension compact, via the presence of two branes which includes the interbrane separation or radion field and the quasi-scalar-tensor theory with specific couplings on both the positive and negative tension branes; see Ref.~\cite{Kanno:2002ia}.

The fact that the gravity sector is stable when the function $f(R,T)$ is separate in the form displayed above, motivates us to investigate new possibilities, for instance, considering $f(R)$ as a Born-Infeld or Gauss-Bonnet term, or even the $f(R,{\cal R})$ term considered before in generalized hybrid metric-Palatini gravity \cite{Rosa:2020uli}, or yet the case of a thick brane in the presence of Lagrange multipliers studied before in \cite{Bazeia:2020jma}. Another line of investigation of current interest is related to the possibility of including another scalar field, changing the field $\chi$ to $\chi_1$ plus $\chi_2$. In this situation, the extra scalar field may be considered to modify the internal structure of the braneworld solutions that we have found in the present work. The new braneworld scenario investigated in this work may also be of current interest from the cosmological point of view. One can study cosmological aspects of braneworlds scenarios with one- and two-brane systems in the presence of bulk scalar fields following the lines of the interesting review in Ref.~\cite{Brax:2004xh}. These issues are presently under consideration, and we hope to report on them in the near future.

\begin{acknowledgments}
JLR is supported by the European Regional Development Fund and the programme Mobilitas Pluss (MOBJD647). MAM and DB acknowledge support from Para\'\i ba State Research Foundation (FAPESQ-PB) grant No. 0015/2019. DB also thanks support from Conselho Nacional de Desenvolvimento Cient\'ifico e Tecnol\'ogico (CNPq), grants Nos. 404913/2018-0 and 303469/2019-6. FSNL acknowledges support from the Fundac\~{a}o para a Ci\^{e}ncia e a 
Tecnologia (FCT) Scientific Employment Stimulus contract with reference CEECINST/00032/2018, 
and funding from the research grants No. UID/FIS/04434/2020, No. PTDC/FIS-OUT/29048/2017 
and No. CERN/FIS-PAR/0037/2019.
\end{acknowledgments}



\begin{thebibliography}{99}

\bibitem{Randall:1999vf}
L.~Randall and R.~Sundrum,
``An Alternative to compactification,''
Phys. Rev. Lett. \textbf{83} (1999), 4690-4693
[arXiv:hep-th/9906064 [hep-th]].

\bibitem{Goldberger:1999uk}
W.~D.~Goldberger and M.~B.~Wise,
``Modulus stabilization with bulk fields,''
Phys. Rev. Lett. \textbf{83} (1999), 4922-4925
[arXiv:hep-ph/9907447 [hep-ph]].

\bibitem{DeWolfe:1999cp}
O.~DeWolfe, D.~Z.~Freedman, S.~S.~Gubser and A.~Karch,
Phys. Rev. D \textbf{62} (2000), 046008
[arXiv:hep-th/9909134 [hep-th]].

\bibitem{Csaki:2000fc}
C.~Csaki, J.~Erlich, T.~J.~Hollowood and Y.~Shirman,
Nucl. Phys. B \textbf{581} (2000), 309-338
[arXiv:hep-th/0001033 [hep-th]].

\bibitem{Harko:2011kv}
T.~Harko, F.~S.~N.~Lobo, S.~Nojiri and S.~D.~Odintsov,
``$f(R,T)$ gravity,''
Phys. Rev. D \textbf{84} (2011), 024020
[arXiv:1104.2669 [gr-qc]].

\bibitem{Bazeia:2015owa} 
  D.~Bazeia, A.~S.~Lob\~ao and R.~Menezes,
  ``Thick brane models in generalized theories of gravity,''
  Phys.\ Lett.\ B {\bf 743}, 98 (2015)
  [arXiv:1502.04757 [hep-th]].


\bibitem{Gu:2016nyo}
B.~M.~Gu, Y.~P.~Zhang, H.~Yu and Y.~X.~Liu,
``Full linear perturbations and localization of gravity on $f(R,T)$ brane,''
Eur. Phys. J. C \textbf{77} (2017) no.2, 115
[arXiv:1606.07169 [hep-th]].

\bibitem{Rohman:2021vvv}
M.~T.~Rohman and Triyanta,
``Localization of scalar field on $f(R,T)$ thick Robertson-Walker brane,''
J. Phys. Conf. Ser. \textbf{1816} (2021) no.1, 012058.

\bibitem{Moraes:2015dee}
P.~H.~R.~S.~Moraes and R.~A.~C.~Correa,
``Braneworld cosmology in $f(R,T)$ gravity,''
Astrophys. Space Sci. \textbf{361} (2016) no.3, 91
[arXiv:1511.08160 [gr-qc]].

\bibitem{Correa:2015qma}
R.~A.~C.~Correa and P.~H.~R.~S.~Moraes,
``Configurational entropy in $f\,(R,T\,)$ brane models,''
Eur. Phys. J. C \textbf{76} (2016) no.2, 100
[arXiv:1509.00732 [hep-th]].

\bibitem{Afonso:2007gc}
V.~I.~Afonso, D.~Bazeia, R.~Menezes and A.~Y.~Petrov,
``f(R)-Brane,''
Phys. Lett. B \textbf{658} (2007), 71-76
[arXiv:0710.3790 [hep-th]].

\bibitem{Zhong:2010ae}
Y.~Zhong, Y.~X.~Liu and K.~Yang,
``Tensor perturbations of $f(R)$-branes,''
Phys. Lett. B \textbf{699} (2011), 398-402
[arXiv:1010.3478 [hep-th]].

\bibitem{Bazeia:2013oha}
D.~Bazeia, R.~Menezes, A.~Y.~Petrov and A.~J.~da Silva,
``On the many-field $f(R)$ brane,''
Phys. Lett. B \textbf{726}, 523-526 (2013)
[arXiv:1306.1847 [hep-th]].

\bibitem{Bazeia:2013uva}
D.~Bazeia, A.~S.~Lob\~ao, Jr., R.~Menezes, A.~Y.~Petrov and A.~J.~da Silva,
``Braneworld solutions for F(R) models with non-constant curvature,''
Phys. Lett. B \textbf{729} (2014), 127-135
[arXiv:1311.6294 [hep-th]].

\bibitem{Gu:2014ssa}
B.~M.~Gu, B.~Guo, H.~Yu and Y.~X.~Liu,
``Tensor perturbations of Palatini $f(\mathcal{R})$-branes,''
Phys. Rev. D \textbf{92} (2015) no.2, 024011
[arXiv:1411.3241 [hep-th]].

\bibitem{Bazeia:2014poa}
D.~Bazeia, L.~Losano, R.~Menezes, G.~J.~Olmo and D.~Rubiera-Garcia,
``Thick brane in $f(R)$ gravity with Palatini dynamics,''
Eur. Phys. J. C \textbf{75} (2015) no.12, 569
[arXiv:1411.0897 [hep-th]].

\bibitem{Bazeia:2015zpa}
D.~Bazeia, L.~Losano, R.~Menezes, G.~J.~Olmo and D.~Rubiera-Garcia,
``Robustness of braneworld scenarios against tensorial perturbations,''
Class. Quant. Grav. \textbf{32} (2015) no.21, 215011
[arXiv:1509.04895 [hep-th]].

\bibitem{daSilva:2017jbx}
P.~M.~L.~T.~da Silva and J.~M.~Hoff da Silva,
``f(R)-Einstein-Palatini formalism and smooth branes,''
Eur. Phys. J. Plus \textbf{132} (2017) no.10, 437.

\bibitem{Gu:2018lub}
B.~M.~Gu, Y.~X.~Liu and Y.~Zhong,
``Stable Palatini $f(\mathcal{R})$ braneworld,''
Phys. Rev. D \textbf{98} (2018) no.2, 024027
[arXiv:1804.00271 [hep-th]].

\bibitem{DeFelice:2010aj}
A.~De Felice and S.~Tsujikawa,
``f(R) theories,''
Living Rev. Rel. \textbf{13} (2010), 3
[arXiv:1002.4928 [gr-qc]].

\bibitem{Nojiri:2010wj}
S.~Nojiri and S.~D.~Odintsov,
``Unified cosmic history in modified gravity: from F(R) theory to Lorentz non-invariant models,''
Phys. Rept. \textbf{505} (2011), 59-144
[arXiv:1011.0544 [gr-qc]].

\bibitem{Cui:2020fiz}
Z.~Q.~Cui, Z.~C.~Lin, J.~J.~Wan, Y.~X.~Liu and L.~Zhao,
``Tensor Perturbations and Thick Branes in Higher-dimensional $f(R)$ Gravity,''
JHEP \textbf{12} (2020), 130
[arXiv:2009.00512 [hep-th]].

\bibitem{Guo:2018tpo}
W.~D.~Guo, Y.~Zhong, K.~Yang, T.~T.~Sui and Y.~X.~Liu,
``Thick brane in mimetic $f(T)$ gravity,''
Phys. Lett. B \textbf{800} (2020), 135099
[arXiv:1805.05650 [hep-th]].

\bibitem{Rosa:2020uli} 
  J.~L.~Rosa, D.~A.~Ferreira, D.~Bazeia and F.~S.~N.~Lobo,
  ``Thick brane structures in generalized hybrid metric-Palatini gravity,''
  Eur.\ Phys.\ J.\ C {\bf 81}, no. 1, 20 (2021)
  [arXiv:2010.10074 [gr-qc]].

\bibitem{Chen:2020zzs} 
  J.~Chen, W.~D.~Guo and Y.~X.~Liu,
  ``Thick branes with inner structure in mimetic $f(R)$ gravity,''
  arXiv:2011.03927 [gr-qc].

\bibitem{Moreira:2021xfe} 
  A.~R.~P.~Moreira, J.~E.~G.~Silva, F.~C.~E.~Lima and C.~A.~S.~Almeida,
  ``Thick brane in $f(T,B)$ gravity,''
  Phys.\ Rev.\ D {\bf 103}, no. 6, 064046 (2021)
  [arXiv:2101.10054 [hep-th]].


\bibitem{Bazeia:2020zut} 
  D.~Bazeia, D.~A.~Ferreira and D.~C.~Moreira,
  ``First order formalism for thick branes in modified gravity with Lagrange multiplier,''
  EPL {\bf 129}, no. 1, 11004 (2020)
  [arXiv:2002.00229 [hep-th]].
  
\bibitem{Xiang:2020qrc}
Q.~Xiang, Y.~Zhong, Q.~Y.~Xie and L.~Zhao,
``Flat and bent branes with inner structure in two-field mimetic gravity,''
[arXiv:2011.10266 [hep-th]].
  
\bibitem{Bazeia:2020jma}
D.~Bazeia, D.~A.~Ferreira, F.~S.~N.~Lobo and J.~L. Rosa,
``Novel modified gravity braneworld configurations with a Lagrange multiplier,''
Eur. Phys. J. Plus \textbf{136} (2021) no.3, 321
[arXiv:2011.06240 [gr-qc]].

\bibitem{Xie:2021ayr}
Q.~Y.~Xie, Q.~M.~Fu, T.~T.~Sui, L.~Zhao and Y.~Zhong,
``First-order formalism and thick branes in mimetic gravity,''
[arXiv:2102.10251 [gr-qc]].

\bibitem{Moreira:2021vcf}
A.~R.~P.~Moreira, J.~E.~G.~Silva and C.~A.~S.~Almeida,
``Fermion localization in braneworld teleparallel f(T, B) gravity,''
Eur. Phys. J. C \textbf{81}, no.4, 298 (2021)
[arXiv:2104.00195 [gr-qc]].


\bibitem{Luis:2021xay}
J.~L. Rosa,
``Junction conditions and thin-shells in perfect-fluid $f\left(R,T\right)$ gravity,''
[arXiv:2103.11698 [gr-qc]].


\bibitem{Kanno:2002ia}
S.~Kanno and J.~Soda,
``Radion and holographic brane gravity,''
Phys. Rev. D \textbf{66} (2002), 083506
[arXiv:hep-th/0207029 [hep-th]].


\bibitem{Brax:2004xh}
P.~Brax, C.~van de Bruck and A.~C.~Davis,
``Brane world cosmology,''
Rept. Prog. Phys. \textbf{67} (2004), 2183-2232
[arXiv:hep-th/0404011 [hep-th]].


\end{thebibliography}
\end{document}